\documentstyle[seceq,epsf]{ptptex}

\def\beq{\begin{equation}}
\def\eeq#1{\label{eq:#1}\end{equation}}
\def\eeqn{\end{equation}}
\def\beqa{\begin{eqnarray}}
\def\eeqa#1{\label{eq:#1}\end{eqnarray}}
\def\eeqan{\end{eqnarray}}
\def\CR{\nonumber \\ }
\def\leqn#1{(\ref{eq:#1})}
\def\bar#1{\overline{#1}}
\def\ee{e^+e^-}
\def\sstw{\sin^2\theta_w}
\def\cstw{\cos^2\theta_w}
\def\Pl{{\hbox{\rm Pl}}} 
\def\N{\hbox{\bf N}}
\def\M{\hbox{\bf M}}
\def\etal{{\it et al.}}
\def\eg{{\it e.g.}}

\newcommand\pubnumber{\large SLAC-PUB-7133}
\newcommand\pubdate{\large April, 1996}

\def\Title#1{\begin{center} {\Large #1 } \end{center}}
\def\Author#1{\begin{center}{ \sc #1} \end{center}}
\def\Address#1{\begin{center}{ \it #1} \end{center}}

\def\doeack{\footnote{Work supported by the Department of Energy,
                     contract DE--AC03--76SF00515.}}
\def\SLAC{Stanford Linear Accelerator Center\\
    Stanford University, Stanford, California 94309 USA}

\newenvironment{Abstract}{\begin{quotation} \begin{center}
                       ABSTRACT
     \end{center}\bigskip  }{\end{quotation}}
\begin{document}
\begin{flushright}\begin{tabular}{l} \pubnumber\\
         \pubdate  \end{tabular}\end{flushright}
\vfill
\Title{The Experimental Investigation of Supersymmetry Breaking}
\vfill
\Author{Michael E. Peskin\doeack}
\Address{\SLAC}

\vfill
\begin{Abstract}
If Nature is supersymmetric at the weak interaction scale, what can we
hope to learn from experiments on supersymmetric particles?  The 
most mysterious aspect of phenomenological supersymmetry
is the mechanism of spontaneous supersymmetry breaking. This mechanism
ties the observable pattern of supersymmetric particle masses to aspects of 
the underlying unified theory 
at very small distance scales.  In this article, I will discuss a systematic
experimental program to determine the mechanism of 
supersymmetry breaking. Both $pp$ and $e^+e^-$ colliders of the next generation
play an essential role.
 
\end{Abstract}
\vfill

\vfill

\begin{center}
  to appear in the proceedings of the Yukawa International Seminar (YKIS-95)\\ 
Kyoto, Japan, August 21-25, 1995 
\end{center}
\vfill


\notypesetlogo  


\markboth{
M. E. Peskin
}{
The Experimental Investigation of Supersymmetry Breaking}

\title{%
The Experimental Investigation of Supersymmetry Breaking}

\author{%
Michael E. {\sc Peskin}\footnote{E-mail address: mpeskin@slac.stanford.edu}
}

\inst{%
Stanford Linear Accelerator Center, Stanford University
\\
Stanford, California 94309 USA}

\recdate{%
April 16, 1996
}

\abst{%
If Nature is supersymmetric at the weak interaction scale, what can we
hope to learn from experiments on supersymmetric particles?  The 
most mysterious aspect of phenomenological supersymmetry
is the mechanism of spontaneous supersymmetry breaking. This mechanism
ties the observable pattern of supersymmetric particle masses to aspects of 
the underlying unified theory 
at very small distance scales.  In this article, I will discuss a systematic
experimental program to determine the mechanism of 
supersymmetry breaking. Both $pp$ and $e^+e^-$ colliders of the next generation
play an essential role.
}

\newpage

\maketitle

\section{Introduction}

Today, many theorists and experimenters expect that, on energy scales
that will soon be probed by accelerators, Nature is supersymmetric, symmetric
between fermions and bosons in the spectrum of fundamental particles.
Part of the motivation for this idea comes from experiment.  
The assumption of supersymmetry allows the values of the standard model
gauge couplings, now precisely measured at LEP and other 
facilities, to be  consistent with grand unification, and it allows the 
observed large value of the top quark mass to lead naturally to electroweak
symmetry breaking. Reviews of these two ideas can be found, respectively, in 
Refs. \citen{landpol}, \citen{rossandi}; more general reviews of 
phenomenological supersymmetry are given in Refs. \citen{nill} and 
\citen{handk}, and \citen{dimopr}.   However, the most compelling arguments
for supersymmetry come from its seductive mathematical beauty
 and its deep connection to string theory and other theories of 
quantum gravity.

The mathematical motivations for supersymmetry lead to research programs that
have very little direct contact with experiment, for example,
investigations of the invariance groups of
supersymmetric theories near the Planck scale, of compactification of 
higher dimensions, and of the nonperturbative, Planck-energy, spectrum.
The intense interest in these topics has led many people to wonder if
the directions of phenomenological and mathematical research in elementary
particle physics have split permanently, so that they can never be rejoined
by future developments. 

 Certainly, the current experimental situation 
offers little evidence that these lines will eventually connect.  This 
will remain true as long as mathematical theory insists that the most
important feature of physics at the TeV energy scale is that it is 
supersymmetric, while experiment sees no direct evidence for physics outside
the standard model. But I often hear much stronger doubts expressed.
Some of my theoretical colleagues argue that,
even in the future, even if
compelling evidence for supersymmetry is discovered, experiment might not
have enough to say that is truly of interest to people making deep 
mathematical investigations.  And my experimental colleauges have been
arguing for years that the 
physics of Planck energies and higher dimensions is so far removed from 
the experimentally accessible domain that it is difficult to conceive of 
any experimental result that would alter or reorient this theoretical 
program.

Personally, I am much more optimistic.  Though I am thoroughly seduced
by the beauty of superstring theory, I recognize that supersymmetry might
not be preserved down to the TeV scale.  However, it is reasonable to 
accept this as a working hypothesis, if only because this provides a 
needed mechanism for electroweak symmetry breaking.  Under this hypothesis,
we should find supersymmetric particles either at the current generation 
of accelerators or at the next step,
 and I intend to keep the faith until a thorough search
is made.  But to answer the questions raised in the previous paragraph,
it is necessary to think through carefully 
to the next step of the experimental program that
will follow this discovery.  Once supersymmetric particles are found,
what can we learn about them?  In particular, can we use them
to gain insight into the truly fundamental issues in Nature which are 
fully revealed only at the unification or superstring mass scale?  I believe
that the answer to this questions is yes, and in this paper I will explain
how it can be done.

Though the grand question of how we imagine the connection of theory and 
experiment is a major issue for our field, and is itself a motivation to 
analyze this problem in advance of the discovery of supersymmetry, there is
another pressing motivation as well.  Experimental high-energy physics cannot
exist without accelerators, and as accelerators become more complex and
expensive, we must be sure that we request the correct ones for the task
at hand.  The time scale for the construction of accelerators is of the 
order of a decade, and in the US we have learned painfully that the process
may conclude with our government's insistence that we scrap everything and
start over from the beginning.  Thus, it is important that we think
clearly about the
physics goals of experiments that will be conducted a decade from now.
For many of the options for the physics of the TeV energy scale, it is very
difficult to see ahead so far.  But the special properties of supersymmetric
theory---the fact that it connects naturally to fundamental theories at
very high energy and also the fact that it involves 
only weak-coupling interactions
at the TeV scale---allow us to look quite far down the road that we will have
to travel experimentally if this hypothesis is realized.  Supersymmetry
thus gives us a concrete example of what the experimental physics of the 
year 2005 could look like, and we can use the detailed picture that it 
provides to draw conclusions about the accelerator facilities we will need
in that era. We must of course keep in 
mind that we do not know what theory of TeV physics Nature actually 
chooses, so that supersymmetry should be only one of many possibilities 
we must survey.  Here, I will only discuss the case of supersymmetry.
A broader survey of models of electroweak symmetry breaking, which reaches
the same general conclusion within a larger set of models, is given 
in Ref.~\citen{meandHitoshi}.

The plan of this paper is as follows:  In Section 2, I will discuss the 
most important open question in supersymmetry phenomenology, that of the
mechanism of supersymmetry breaking.  I will review the general constraints
on the nature of supersymmetry breaking and the various models of this 
phenomenon which have been proposed in the literature.  In Section 3, I 
will present a three-step experimental program to distinguish the various
models discussed in Section 2 and thus resolve experimentally the origin
of supersymmetry breaking.  In Section 4, I will discuss the contributions
to this program which can be expected from $pp$ colliders, and, in particular,
from the LHC.  In Section 5, I will discuss the contributions to this 
program which can be supplied by $\ee$ linear colliders of the next generation,
and I will demonstrate the essential role that $\ee$ experiments 
will play in 
this investigation.  Section 6 will give some general conclusions.

\section{Phenomenological Theory of Supersymmetry Breaking}

In this paper, I will assume that  Nature is supersymmetric at the 
weak interaction scale, and that ingredients of the supersymmetric
theory provide the mechanism of electroweak symmetry breaking.
This hypothesis can be used to place upper limits on the masses
of supersymmetric particles.
If supersymmetry is the mechanism of electroweak symmetry breaking,
 the $W$ and $Z$ masses are directly related to the mass scale of 
supersymmetry.  The exact correspondence between the $Z$ mass and the 
mass of, say, the supersymmetry partners of the electron is model-dependent.
However, the ratio between these masses cannot be made arbitrarily large
without a fine adjustment of parameters.  Several groups have tried to 
characterize the reasonable range of allowable fine tuning and to convert
this range to a set of bounds on particle masses.\cite{nat}  The most
 characteristic of these are limits on the masses of the $W$ and 
gluon partners,
\beq 
             m(\widetilde w) < 250 \hbox{\rm GeV} \ , \qquad
             m(\widetilde g) < 800  \hbox{\rm GeV} \ . 
\eeq{natlim}
These limits imply that supersymmetric partners should be found, at the 
latest, at the LHC and at the next generation of $\ee$ colliders.

In my analysis here, I will assume that the first signals of supersymmetry
have been found, and I will be concerned with the questions at the 
 next level to be explored.  When I discuss the results of specific 
experiments, I will make one further assumption, that 
there is a conserved `R-parity' which makes the lightest
supersymmetric partner stable.
Then supersymmetric particle production will be 
characterized by signatures of missing energy and unbalanced
momentum which should be visible both at lepton and at hadron 
colliders. If supersymmetric particles are present but R-parity is
violated, there should be a similarly rich program involving signatures
with lepton or baryon number violation; see, for 
example, Refs. \citen{handk}  and \citen{Rviol}.

Assuming, then, that supersymmetric partners are found,
what is the next question that we would like to answer?
 A simple reply is that we will want
to measure the masses of these supersymmetric partners and understand their
properties systematically.  I will
address this issue in more detail.

The equation of motion of a supersymmetric extension of the standard model
has three parts.  Of these, two are highly constrained by supersymmetry:
The gauge interactions of superpartners are fixed by their $SU(3)\times 
SU(2)\times U(1)$ quantum numbers, and the renomalizable couplings of 
quark, lepton, and Higgs partners are fixed to be equal to the corresponding
couplings of the standard model.  However, the third piece of the puzzle is
a complete mystery.  If we wish to understand why the partners of quarks and
leptons are heavy, we must appeal to some
mechanism of spontaneous
supersymmetry breaking.  This mechanism is unknown and
is not constrained by a direct connection to any known physics.  This 
mechanism controls the regularities of the supersymmetric mass spectrum
and the possible mixings between superpartner states.  It also controls 
the other important qualitative features of the theory.  For example,
the various sources of the Higgs boson masses which lead eventually to 
$SU(2)\times U(1)$ breaking have their origin in supersymmetry breaking.

Supersymmetry breaking also connects the phenomenology of supersymmetry to
the truly deep questions about the structure of elementary particles.  If 
Nature is supersymmetric and weakly-coupled at the TeV scale, it is 
reasonable that the strong, weak, and electromagnetic interactions 
are grand-unified at very high energy.  We already have some information
about this unification from the values of the gauge couplings, and from 
the ratio $m_b/m_\tau$.  From a first point of view, the measurement of
supersymmetry breaking will give us access to a new set of 
parameters, outside
those which are directly connected to the standard model couplings,
from which we can obtain new pieces of evidence on the properties of the
unified theory.  But there is also a 
more ambitious point view.  In models in which one attempts to derive the 
whole structure of Nature from a superstring model, supersymmetry breaking
typically arises from sectors of the theory which have an essential 
connection to the string origins of the model, for example, from the 
interactions of the `dilaton' and `moduli' fields and the gauge fields of 
`hidden' sectors,\cite{janandv,iban} or from the transmutation of 
dimensions that occurs in string theory at strong coupling.\cite{witten}
If we can study the pattern of supersymmetry breaking experimentally, we
might obtain a direct window into these deep structures.

\subsection{Issues of supersymmetry breaking}

What, exactly, do we wish to know about supersymmetry breaking? At the first
level of any discussion of the physics of supersymmetry breaking, 
two questions arise. The answers to these questions would take us a long
way toward an understanding of supersymmetry breaking and its
 relation to the other fundamental interactions.

The first of these questions
 is the mass scale of supersymmetry breaking and, as a 
closely connected issue, the scale of the transmission of supersymmetry
breaking.  The interplay of these scales deserves some explanation. 
To begin, we should recall why it is that the quarks and leptons are
expected to be lighter than their superpartners, rather than
the other way around.  Quarks and leptons can receive mass only if 
$SU(2)\times U(1)$ is spontaneously broken.  
However, their partners---squarks and sleptons---are
scalars, and there is no principle of quantum field theory that
prohibits scalar fields from obtaining a mass.	What keeps the quark and 
lepton partners light is their supersymmetry relation to their fermion
partners.  If supersymmetry is spontaneously
broken in some sector  of Nature, and this sector communicates with the 
quarks and leptons and their partners through some interactions, the 
supersymmetry breaking will be transmitted to the 
squarks and sleptons
to produce scalar masses and other soft interactions.  Call the scale of 
these masses $m_S$.  The Higgs boson masses will also be of scale $m_S$.
The Higgs vacuum expectation value will also be of size $m_S$, up to 
coupling constants, and so $m_S$ will determine the location of the weak
interaction scale.  Then, finally, masses are fed down to the quarks and 
leptons according to the strength of their coupling to the Higgs sector.

The value of $m_S$ is determined by the underlying physics responsible for
spontaneous supersymmetry breaking.  Let $\Lambda$ be 
the mass scale of spontaneous supersymmetry breaking, and let $M$ be the 
mass of the particles that connect the symmetry-breaking sector to the 
quarks, leptons, and standard model gauge bosons.  I will refer to $M$
as the `messenger scale', and it will play a crucial role in our analysis.
Though the relation between $M$, $\Lambda$, and $m_S$ is model-dependent,
the general form of this relation is given by 
the equation
\beq
    m_S \sim    {\Lambda^2\over M} \ ,  
\eeq{mSis}
so that different choices for $\Lambda$ and $M$ are correlated by the fact
that they must generate $m_S \sim m_Z$.

By default, gravity (or supergravity) is the messenger.  This was made clear
in the beautiful foundational papers of Cremmer and 
collaborators,\cite{CFGvP} 
who showed explicitly how supersymmetry breaking is transferred
from the original symmetry-breaking sector to the rest of Nature through
supergravity interactions.  More generally, the messenger interactions might
be associated with the grand unified scale or other flavor physics, with 
some intermediate scale, or with the standard model gauge interactions.
The nature of the messenger plays an important role in determining the 
form and selection rules for the supersymmetry breaking masses and 
interactions.

If this were our only information about $M$ and $\Lambda$, there would be 
considerable room for speculation.  Fortunately, the range of possible 
theories that lead to $M$ and $\Lambda$ is limited by additional constraints.
These stem from
the second problem
that the mechanism of supersymmetry breaking must solve, the `supersymmetric
flavor problem'.  To understand this issue, let us write the formula for
the mass matrix of the scalar partners of the $d$, $s$, $b$ quarks.  Since
in supersymmetry left- and right-handed fermions have independent complex
scalar fields as their superpartners, I will write this matrix as a 
$2\times 2$ matrix of $3\times 3$ blocks, acting on a vector
\beq 
 \pmatrix{{\widetilde d}^i_L \cr {\widetilde d}^i_R\cr}
\eeq{dvector}
where $i$ is the generation label, $i=1,2,3$.  The mass matrix gets 
contributions from four sources, two of which are supersymmetric---the 
quark mass matrix $m_d$ of the 
standard model, and the combination of this term with the Higgs mass
parameter $\mu$---and two of which arise from supersymmetry-breaking---scalar
field mass matrices $m^2_{dL}$ and $m^2_{dR}$, and a mixing term generated 
by a supersymmetry-breaking 3-scalar term involving the Higgs field.  The 
final result is a matrix
\beq
   {\cal M^2} = \pmatrix{ m^2_{dL} + m_d m_d^\dagger & 
                         - m_d (A + \mu \tan\beta)\cr
                           - m_d^\dagger (A + \mu \tan\beta)&
                   m^2_{dR} + m_d^\dagger m_d\cr} \ .         
\eeq{dmasses}
In this equation, $\tan\beta$ is the ratio of the two Higgs field vacuum
expectation values required in the minimal supersymmetric extension of the
standard model: $\tan\beta = \hbox{$<h_2^0>/<h_1^0>$} = 
v_2/v_1$. This parameter infects all of 
supersymmetry phenomenology.  I have simplified the expression by writing the
3-scalar term in terms of a constant parameter $A$.  In principle, this 
could also be a matrix with flavor indices.
                    
The quark mass matrix $m_d$ is not intrinsically diagonal.  In standard
weak interaction phenomenology, we
diagonalize it with matrices $V_L$ and $V_R$ which eventually 
become  ingredients of the Cabbibo-Kobayashi-Maskawa mixing matrix:
\beq
  m_d =  V_L^\dagger \pmatrix{ m_d & & \cr & m_s& \cr & & m_b\cr} V_R \ .
\eeq{ddiag}
Then the $Z^0$ couplings are automatically flavor-diagonal;
flavor-changing neutral current effects appear only in loop diagrams, and
only proportional to products of quark mass differences.  This lead to the 
observed suppression of flavor-changing neutral current processes.  However,
 the mass
matrix  \leqn{dmasses}
contains new sources of flavor violation through the 
supersymmetry-breaking scalar mass matrices $m^2_{dL}$, $m^2_{dR}$.
Unless the diagonalization of $m_d$ also diagonalizes these matrices,
diagrams with supersymmetric particles in loops can provide new and 
dangerous sources of flavor violation.	For example, applying this logic
to the contribution to the $K_L$--$K_S$ mass difference due to gluino 
exchange, Gabbiani and Masiero\cite{gandm} have derived the bound
\beq
 {(V_R m^2_{dR} V_R^\dagger)_{12} \over m^2_{\widetilde d}} < 
         10^{-2} \bigl( {m_{\widetilde d}\over 300 \ \hbox{\rm GeV}}\bigr)^2
                 \ .   
\eeq{Vbound}
Similar bounds on the flavor violation of the supersymmetry-breaking mass
terms have been discussed by many authors.\footnote{See, for example,
Ref. \citen{nill}; some recent articles are given 
in Ref. \citen{moreflavor}.}

\subsection{Models of supersymmetry breaking}

Why should the supersymmetry-breaking scalar masses be diagonal in the  
same basis as the standard model mass terms?  There are a large number of 
explanations for this in the literature.  These explanations divide into
general classes which express the range of possibilities for the
underlying physics of supersymmetry breaking.  On the one hand, it is 
possible that the supersymmetry breaking scalar masses  are universal 
among generations, so that the mass matrices $m_{dL}^2$, $m^2_{dR}$ are 
proportional to 1 and thus diagonal in any basis.  Or these mass matrices
may have structure, but they might also have a reason to be diagonal in	
the basis set by the mass matrix.  On the other hand, the mechanism for
the specific form of these matrices might be predetermined by the 
short-distance physics, or it might arise as the result of dynamical 
effects on larger scales.  Thus, we have four classes of models:

\medskip
\begin{enumerate}

\item  {\bf Preset Universality.} This is the original schema for supersymmetry
model building which was proposed in the early papers of Dimopoulos and 
Georgi\cite{dandg} and Sakai.\cite{sakai}
  It is realized elegantly, with $M$ equal to the Planck
mass  $m_\Pl$,
in models in which supersymmetry is broken at a high scale and
the breaking is communicated by supergravity.\cite{ACN}
  Other agents which couple
universally to quarks and leptons can also give models of this structure.

\item {\bf Dynamical Universality.} This class encompasses a broad range of 
models in which the supersymmetry-breaking mass matrices are fixed in a 
manner determined only by the standard model gauge couplings of superpartners.
It includes the `no-scale' models in which $m^2_L$, $m^2_R$ are 
zero at the fundamental scale and are generated by radiative 
corrections,\cite{enandk}
a model of Lanzagorta and Ross\cite{ross}
in which  $m^2_L$, $m^2_R$ are determined by an infrared
fixed point, and models studied by Dine, Nelson, Nir, and 
Shirman\cite{dns}
 in which
supersymmetry is broken at a low scale and communicated through the 
standard model gauge interactions.

\item  {\bf Preset Alignment.}  This class of models attempts to build up
      the supersymmetry breaking mass matrices using the same principles
 that one uses to construct the standard model quark mass matrices
      (for example, the successive breaking of discrete symmetries).
   Then these symmetry principles can insure that the two sets of matrices
 are diagonal in the same basis, without flavor-degeneracy of scalar 
     masses.\cite{nls}
 In this class of models, it is natural for the messenger
      scale to be of the order of the grand unification scale.

\item  {\bf Dynamical Alignment.}  In this class of models, the relative 
       orientation of the supersymmetry-breaking and standard model mass
       matrices is a free parameter in the underlying theory and is 
       determined to be aligned
by radiative corrections.  The one current example of a 
      model in this class has $M$ near the Planck scale,\cite{dandgiud},
     leading to a phenomenology very similar to that of the class just
 above.  A  very
        low value of $M$ might be  more natural in this scheme and may
       lead to some different options.

\end{enumerate}
\medskip

Here are four broad classes of possibilities for the mechanism of 
supersymmetry breaking.  It is interesting to lay out the various
possibilities in this way, because it makes clear that every specific 
solution to the supersymmetric flavor problem entails a choice of $M$ and
therefore of $\Lambda$.  If we can recognize experimentally which possibility
Nature chooses, we can also infer the nature of the messenger and perhaps
the specific origin of supersymmetry breaking.

\section{An Experimental Program}

How can we decide which mechanism is chosen?  At a certain  level, it is 
obvious that the answer can be found by measuring the spectrum of 
superparticle masses.  To probe more deeply, we should ask which of these 
measurements are easy and which are very challenging, and whether the 
measurements that are reasonably straightforward
can actually give us the information we are looking for.

To understand how we will learn about these fundamental issues from 
measurements, it is necessary to work out the correspondence between
properties of the supersymmetry spectrum and the various hypotheses
described above.  I will descibe that correspondence in Section 3.3.
To prepare the way, we must first discuss two issues that provide the 
baseline for that analysis, the masses of the gauge boson superpartners
and the value of the Higgs sector parameter $\tan \beta$.

\subsection{Gaugino masses}

\medskip
In the discussion of the previous section, we concentrated our attention on
the masses of the scalar partners of the quarks and leptons. The masses
of the fermionic partners of gauge bosons---gauginos---did not seem to play
an important role.  But in fact, a precise understanding of gauginos is a 
prerequisite to any detailed exploration of supersymmetry.  This is true 
for two reasons.  First, gaugino masses influence scalar masses through
radiative corrections. Second, the nature of the gaugino mass matrix affects
the general phenomenology of supersymmetry, as viewed by collider 
experiments.  In this section, I will review both of these issues.

\begin{figure}
 \epsfysize=1.0in
\centerline{\epsffile{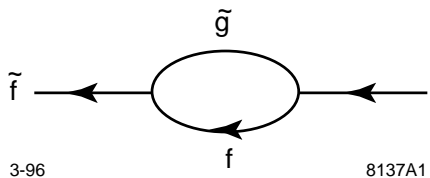}}
\caption{ Feed-down of gaugino masses into scalar masses.}
\label{fig:Feed}
\end{figure}

The systematics of gaugino masses forms an essential part of the scalar 
mass problem due to the diagram shown in Fig. \ref{fig:Feed}. 
The scalar masses are
renormalized, as shown,  by the transition 
to a gaugino and a quark or lepton.
This process gives a correction to the mass which is described by the
renormalization group equation
\beq
 -{d m_f^2\over d \log Q} = \sum_i {2\over \pi} C_i \alpha_i m_i^2 \ ,
\eeq{grenorm}
where $m_f^2$ are the scalar masses and $m_i$ are the gaugino mass parameters
generated by supersymmetry breaking. 
The coupling constants  $\alpha_i$
are the standard model couplings, evaluated at the weak interaction 
scale,
 normalized as in grand unification:
$\alpha_3 = \alpha_s$, $\alpha_2 = \alpha_w$, $\alpha_1 = (5/3)\alpha'$.
The $C_i$ are Casimir coefficients:
\beq
  C_1 = {3\over 5}Y^2\ , \qquad C_2 = \cases{{3\over 4} & L \cr 
               0 & R\cr} \ , 
\qquad C_3 = \cases{0 & $\ell$ \cr {4\over 3} & q\cr}
                 \ .    
\eeq{Cdefs}
The renormalization group equation must be integrated from the messenger 
scale $M$ to the weak scale.  One of the ways we can determine the value of
the messenger scale is to estimate how far this renormalization group equation
has been evolved in order to produce the observed spectrum of scalar 
masses.  To do that, we require the values of the gaugino masses, to set the
overall scale of this effect.

At the same time that they renomalize the scalar masses, the gaugino masses
evolve by their own renormalization.  This means that a simple spectrum
of gauginos at one scale will acquire structure as we move to a different
scale.  The simplest possible picture of gaugino masses is that they are 
grand-unified, that is, they are all equal at the grand unification scale.
From this starting point, the one-loop renormalization group equation gives
an interesting pattern at lower scales:  The gaugino masses evolve so as to 
remain proportional to the gauge couplings:
\beq
  {m_1\over \alpha_1} = {m_2\over \alpha_2} = {m_3\over \alpha_3}
                             \ .    
\eeq{gunif}
I will refer to this systematic relation as `gaugino unification'. The 
simple relation is corrected by the two-loop terms in the 
renormalization group equations and by finite one-loop corrections at the
weak scale.\cite{twoloop}
  The only large correction comes in the finite contributions
which relate the short-distance gluino mass to the physical gluino 
mass,\cite{glmassD}
a problem reminiscent of the problems of the quark mass definition in QCD.

It is interesting to ask how broad a class of models obey gaugino unification.
Obviously, if there is no grand unification, there is no reason for this 
relation to be true.  However, one of the phenomenological virtues of 
supersymmetry is that it allows the grand unification of couplings, and so
it is reasonable to assume this in model-building.  Still, grand unification
does not necessarily imply gaugino unification.  On one hand, the messenger
scale might be well below the grand unification scale, so that the 
physics of gaugino mass generation is not grand unified.	
On the other hand,
it is possible that the field which breaks supersymmetry is not a singlet
of the grand unification group.
A violation of gaugino universality would be a signal of one of these
mechanisms and thus would be of great experimental importance.
Curiously, though, the
simplest models of each type actually respect the relation \leqn{gunif}.
For example, in the model of Ref. \citen{dns}, supersymmetry is communicated
from a unique standard model singlet field to a vectorlike multiplet of fields
and from there though standard model gauge interactions to the partners of
gauge bosons.  The messenger scale is well below the grand unification scale,
but the vectorlike multiplet must be a complete $SU(5)$ representation
(\eg, $(5 + \bar 5)$) so as not to spoil the unification relation of the 
low energy gauge couplings.  These assumptions
imply that the gaugino masses are related by \leqn{gunif}.
Similarly, if supersymmetry breaking is communicated at the  grand unification
or scale, the communication of supersymmetry breaking by an 
$SU(5)$-nonsinglet field involves a nonrenormalizable interaction or a 
perturbative correction and thus would be suppressed with respect to any 
nonzero contribution from singlet fields.\cite{tsuka}

The experimental measurement of the gaugino mass
 parameters $m_i$ brings in some additional issues which we might call 
problems of supersymmetry engineering.  These are not problems of 
principle, but they must be resolved to understand the deeper aspects of 
supersymmetry phenomenology. 

 The parameter $m_3$ is the only contribution to the 
mass of the supersymmetry partner of the gluon, the gluino, up to the 
usual problems of defining the mass of a colored particle.  I will discuss
techniques for the  measurement of the gluino mass in Section 4.  
For the supersymmetry partners of $W$, $Z$, and $\gamma$, however, there 
are additional effect that contribute to their masses.  Even in a 
supersymmetric situation, the partners of $W$ and $Z$ will obtain mass
from the Higgs mechanism.  This mass term couples the fermionic parters of 
the vector bosons to the fermionic partners of the Higgs bosons.  These 
latter particles can obtain mass also from a supersymmetric mass term
$\mu$, and we know from the non-observation of light superpartners at the 
$Z^0$ that $\mu$ is nonzero.\footnote{A tiny corner of parameter space is 
still available; see Ref. \citen{FPT}.}

These effects are summarized as a mixing problem involving the vector boson
and Higgs boson superpartners.  Supersymmetric models necessarily include
two Higgs doublets $h_1$, $h_2$;
therefore, they contain physical charged Higgs fields, 
which have fermionic partners.  Denote the left-handed fermion partners of
$W^+$ and $h_2^+$ by $\widetilde w^+$, $\widetilde h^+_2$, and adopt a 
similar notation for the left-handed fermion partners of $W^-$ and $h_1^-$.
Then the charged fermionic superparticles have a mass matrix, including all
three of the effects described in the previous paragraph, which takes the
form
\beq
 \pmatrix{-i \widetilde w^- & \widetilde h_1^- \cr}
       \pmatrix{m_2  & \sqrt{2} m_W \sin\beta \cr 
       \sqrt{2} m_W \cos\beta  & \mu \cr}
 \pmatrix{-i \widetilde w^+ \cr \widetilde h_2^+ \cr} \ .
\eeq{charginom}
This mass matrix is asymmetric, and its diagonalization will generally
 require a different mixing angle for the positively and negatively charged
left-handed fermions.  In a similar way, the partners of the $Z$, photon, and neutral 
Higgs bosons have a $4\times 4$ mixing problem: 
\beq
       \pmatrix{m_1 & 0& -m_Z\sin\theta_w\cos\beta &
   m_Z\sin\theta_w\sin\beta  \cr
          0   &m_2  & m_Z\cos\theta_w\cos\beta &
 -m_Z\cos\theta_w\sin\beta \cr 
     -m_Z\sin\theta_w\cos\beta &
 m_Z\cos\theta_w\cos\beta  & 0& -\mu \cr     m_Z\sin\theta_w\sin\beta &
 -m_Z\cos\theta_w\sin\beta & -\mu & 0 \cr}
\eeq{neutralm}
acting on the vector $(-i \widetilde b, -i \widetilde w^3, \widetilde h_1^0,
 \widetilde h_2^0)$.
The mass eigenstates of \leqn{charginom} and \leqn{neutralm}
are called, respectively, `charginos' and
`neutralinos' and are denoted $\widetilde\chi_i^+$,  $\widetilde\chi_i^0$.

\begin{figure}
 \epsfysize=3.3in
\centerline{\epsffile{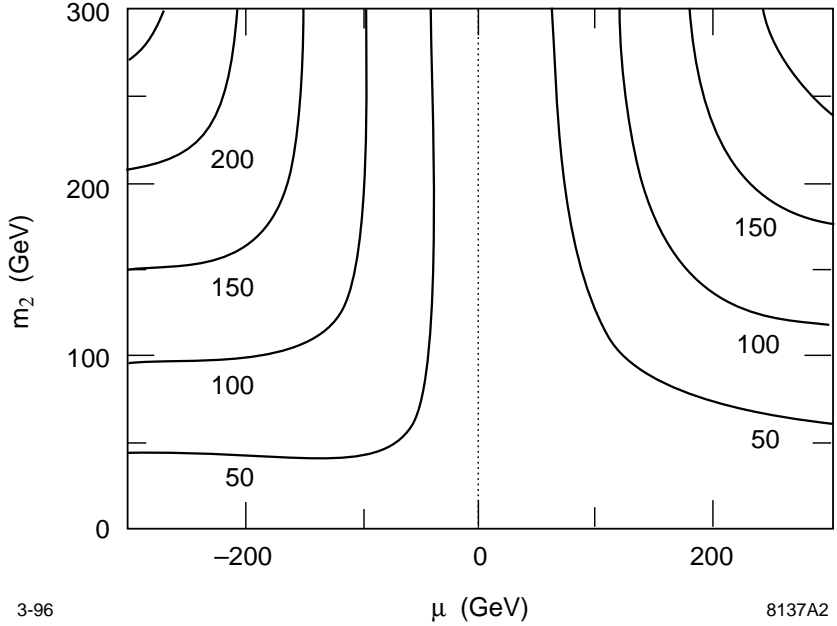}}
\caption{ Lines of constant $\widetilde\chi_1^+$ mass
    in the $(m_2,\mu)$ plane, for $\tan\beta=4$.}
\label{fig:CMasses}
\end{figure}

One cannot, then, extract $m_1$ and $m_2$ simply by observing the masses of 
supersymmetric particles.  It is also necessary to understand which values of
the mixing angles Nature has chosen.  Constraints 
coming from searches for charginos and neutralinos are often plotted on 
the plane of $m_2$ versus $\mu$, at a constant value of $\tan\beta$.  The
lines in this plane representing
constant mass of the lighter chargino, for $\tan\beta = 4$, are 
shown in Fig. \ref{fig:CMasses}. 
 Toward the bottom of this figure, the masses of the 
lightest charginos and neutralinos are close to $m_2$ and $m_1$, and these
particles are composed dominantly of the gauge partners.  Toward the top 
of the figure, the lightest chargino and neutralino become degenerate at the
value $\mu$ and behave like the partners of Higgs bosons.  This means that 
it is essential, both for the extraction of the supersymmetry breaking 
parameters and for the more general understanding of the signatures of 
supersymmetry that experiments should determine where we actually sit
in the $(m_2,\mu)$ plane.

\subsection{$\tan \beta$}

The set of parameters needed for a precise understanding of the spectrum
of superpartners also includes $\tan\beta = v_2/v_1$, the 
ratio of the Higgs field
vacumm expectation values. We have already seen that $\tan\beta$ 
appears as a parameter in the gaugino mixing problem.  This parameter
also plays a role in the formula for the scalar masses.  Through the 
supersymmetrization of the gauge interactions, all quark and lepton 
partners receive a `D-term'
contribution to their masses of the form
\beq
       \Delta m^2_D = - m_Z^2 \bigl({\tan^2\beta-1\over \tan^2\beta + 1}\bigr)
             (I^3 - Q \sstw) \ ,
\eeq{Dterm}
where $I^3$ and $Q$ are the electroweak quantum numbers.\footnote{If the 
theory contains additional gauge bosons, there are additional D terms. I
include these in the model-dependent part of the scalar masses.}

 More generally, 
any discussion of the experimental signatures of supersymmetry brings
in many sources of dependence on $\tan\beta$, through the production and 
decay amplitudes for gauginos and Higgs bosons.  Thus, it is 
important to find a model-independent method for determining this 
paramter.

\subsection{Scalar partner masses}

Once the gaugino mass parameters and
$\tan\beta$ are determined experimentally, we will have established a 
proper foundation for a discussion of the 
spectrum of scalar masses. I will now discuss how the scalar masses can
be analyzed, 
and what variety of patterns the various models of Section 2.2
produce.

In general, the formula for a scalar partner mass has three components.
First, there is the underlying supersymmetry-breaking mass term.  At least
for the light generations, for which we can ignore the Yukawa couplings to 
the Higgs sector, this term is not renormalized at the level
of one-loop renormalization group equations.  Second, there is the 
contribution fed down from the gaugino masses, obtained by integrating
the renormalization group equation \leqn{grenorm}.  Finally, there is the 
D-term contribution \leqn{Dterm}.  Once $\tan\beta$ is known, this last
contribution can be computed in a model-independent way and subtracted;
I will define the reduced scalar partner masses
\beq
          \bar m^2_f = m^2_f - \Delta m^2_D(I^3,Q) \ .
\eeq{barmdef}

Next, we must deal with the mass
contribution due to gauginos.  The result of integrating \leqn{grenorm} 
can be conveniently written
\beq
   \bar m^2_{f}= \bar m_{f0}^2 + 
\Bigl( \sum_i 2 C_i {\alpha_i^2 - \alpha_{iM}^2
         \over b_i \alpha_2^2} \Bigr) \cdot m_2^2 \ .
\eeq{gaugefeed}
In this equation, $i = 1,2,3$ runs over the standard model gauge groups.
The $C_i$ are the Casimirs from \leqn{Cdefs}.  The $b_i$ are the
 renormalization 
group coefficients; these are given by 
$b_i = (-33/5,-1,3)$ for $i = 1,2,3$
 in minimal supersymmetry.  Finally, the
$\alpha_{iM}$ are the values of the coupling constants at the messenger
scale $M$.  In writing this equation, I have assumed gaugino universality to 
convert the gaugino masses to the single value $m_2$, which should be 
precisely known.  I emphasize again that gaugino universality is an 
assumption, but one which can be confirmed or refuted experimentally as 
part of the broader exploration of supersymmetry.

In Section 2.2, the class of models exhibiting dynamical universality
included models in which the messengers of supersymmetry breaking were the
standard model gauge interactions.  In models of this type, gaugino masses
are generated directly by one-loop diagrams involving the
 supersymmetry breaking sector, and 
scalar masses are generated at the two-loop level.  I have already noted
that these models can naturally lead to gaugino unification.  A particular
model of Dine, Nelson, Nir, and Shirman\cite{dns} gives also gives a simple
spectrum of scalar masses:
\beq
   \bar m^2 = \Bigl( \sum_i 2 C_i {\alpha_i^2
         \over \alpha_2^2} \Bigr) \cdot m_2^2 \ ;
\eeq{gaugetwo}
in this formula, the coefficient 2 depends on the model assumptions, while
the general structure is characteristic of this mechanism for the 
communication of supersymmetry breaking.	

The simplicity of the formula \leqn{gaugetwo} and its curious 
resemblance to \leqn{gaugefeed} motivates us to consider the following
device for exhibiting the spectrum of quark and lepton superpartners.
We plot the ratio $\bar m/m_2$ against a weighted combination of 
Casimirs,
\beq 
  C = \bigl( \sum_i C_i {\alpha^2_i\over \alpha^2_2} \bigr)^{1/2}\ .
\eeq{Cfactor}
The prediction of \leqn{gaugetwo} is that the superpartner spectrum 
is a straight line on this plot.  Thus it is reasonable to call this 
device the `Dine-Nelson plot'.

\begin{figure}
 \epsfysize=5.2in
\centerline{\epsffile{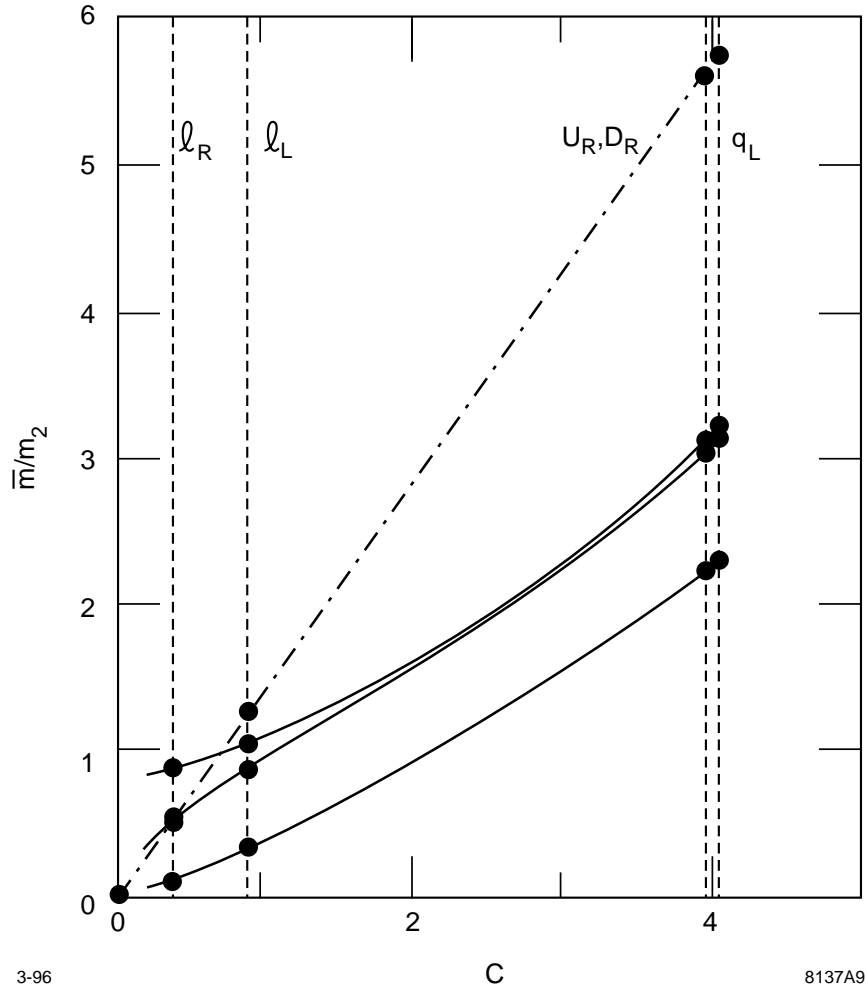}}
\caption{Some reference models displayed on the 
       Dine-Nelson plot.  The solid lines show the integration of the
       renormalization group equation for two values of the messenger
       scale. The dotted line shows the linear relation predicted in the
         model of Dine, Nelson, Nir, and Shirman.}
\label{fig:RefsDN}
\end{figure}

Models in which the scalar masses 
come dominantly from the renormalization group effect \leqn{gaugefeed}
also have a relatively simple form on the Dine-Nelson plot.  
In Fig. \ref{fig:RefsDN},
I have plotted the contributions from renormalization-group running
for three values of the messenger scale---a low scale $M = 100$ TeV, the
grand unification scale $2\times 10^{16}$ GeV, and the fundamental 
scale of superstring theory, $10^{18}$ GeV.
  As a comparison, I have also plotted
the result \leqn{gaugetwo}.  It is important to note that the Casimir
$C$ is not continuously variable but rather takes only fives distinct
values, those for the $SU(2)\times U(1)$ multiplets of the standard model,
$\ell_R$, $L_L$, $d_R$, $u_R$, and $Q_L$. Of these, the values of 
$C$ for $d_R$ and $u_R$ (and  also the gaugino contributions
to their scalar masses) are highly degenerate.  So the Dine-Nelson plot 
is really defined by the value of the masses at these specific
points.  The curves
in Fig. \ref{fig:RefsDN} are intended only to guide the eye.

\begin{figure}
 \epsfysize=6.7in
\centerline{\epsffile{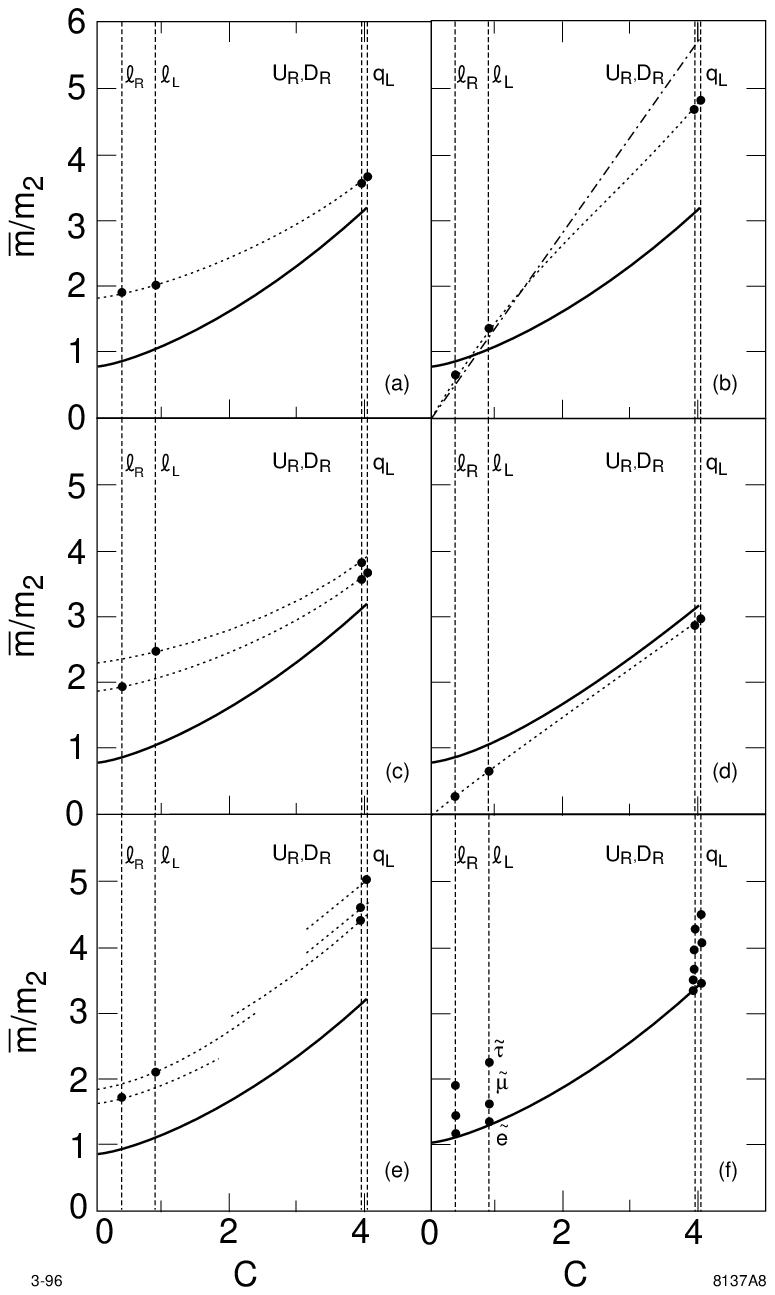}}
\caption{ Six classes of models of supersymmetry breaking,
       displayed as patterns on the 
       Dine-Nelson plot.  The solid reference line is the result of 
       integrating the renormalization group equation from the string 
        scale.  The models (a)-(f) are described in the text.}
\label{fig:ClDN}
\end{figure}

The device of the Dine-Nelson plot gives us a concrete way to view the 
distinctions between the various classes of models reviewed in Section 2.
In Fig. \ref{fig:ClDN},
 I have plotted the spectrum of quark and lepton partners for
each of six representative models.  The values of the masses are compared
to the position of the top solid curve from
 Fig. \ref{fig:RefsDN}, representing the
gaugino loop contribution of Fig. \ref{fig:Feed} integrated from the 
superstring scale.  In supersymmetry phenomenology, the 
squarks and sleptons of the third generation
are often split from the first two due to their coupling to the 
particles of the Higgs sector.  I will ignore that complication here.

Model (a) is  a typical model with preset universal scalar masses 
communicated by supergravity.\cite{ACN}
 The values of the masses sit a constant
distance in (mass)$^2$ above the solid curve.  This increment is 
positive, so that the theory does not develop an instability at the 
fundamental scale.\footnote{There are realistic models which avoid
this constraint in which our vacuum is not the global minimum of the
potential but is stable over cosmological time; see
Ref. \citen{kls}.}
 Notice that the sleptons
are typically lighter than the squarks, but the ratio of these masses depends
on the size of the original supersymmetry-breaking mass term relative to 
that generated by renormalization group corrections.

Model (b) is the Dine-Nelson-Nir-Shirman model.\cite{dns}
  I have made some small 
improvements of the formula \leqn{gaugetwo}, evaluating the coupling constants
at a more realistic scale of about 100 TeV, and then adding the 
renormalization group enhancements as the masses come down to the weak 
interaction scale.  The dashed line is copied from Fig. \ref{fig:RefsDN}.
Notice that in  this class of models the slepton
masses are rather small, and also different by a factor 2 between 
the partners of left- and right-handed leptons.

Model (c) is a variant of the supergravity models which has been considered 
in Ref. \citen{candh}.
 Here the original supersymmetry-breaking scalar masses
are universal among generations for a given set of gauge 
quantum numbers.  However, the values of these masses depend on the 
quantum numbers, for example, differing for the particles that belong to 
10 and $\bar{5}$ representations of $SU(5)$.  Models in which 
there are large contributions to the scalar masses from new D terms
due to extended gauge interactions, as in Refs. \citen{kawa} and 
\citen{koldam},
 and the superstring-based models of Ref. \citen{lopeznk}, generate patterns
similar to these.

Model (d) is a model with dynamical universality presented by Choi.\cite{choi}
 In this model, the original supersymmetry breaking masses are zero, so that
the final masses are determined only by the renormalization group effect,
as in `no-scale' models.  However, for Choi, the messenger scale is 
$F_a$, the axion decay constant, and the messenger interactions are those
associated with Peccei-Quinn symmetry breaking.

Model (e) illustrates an idea for dynamical universality due to 
Lanzagorta and Ross.\cite{ross}  In 
this model, the supersymmetry-breaking masses are driven to the fixed points
of the renormalization group equations for a more complex underlying theory
at a high scale.  The locations of the fixed points depend on the 
standard model quantum numbers of the quark and lepton partners, but not on 
the generation.  In principle, the pattern of soft masses is predicted
by the underlying model.

Model (f) is an example of a model with preset alignment.  In such a scheme,
     the three supersymmetry-breaking mass parameters for each set of 
      standard-model quantum numbers are distinctly different.
     Though this is not required in these models, I have drawn the figure
to suggest that the masses, for each set of quantum numbers, have an 
      asymptote which is the solid line; this would suggest that the 
       messenger scale is the Planck or string scale, and that the discrete
      symmetries which regulate the alignment of the mass matrices are 
     characteristic of superstring or other deep-level physics.

Each one of the possibilities presented here is interesting as a plausible 
model of the origin of supersymmetry breaking.  The range of possibilities
is fun to think about, and is certainly not exhausted by these cases.  That
there should be such a wide range is no surprise.  In physics, every time
we open another door to speculation, manifold possibilities are revealed, and
the one chosen by experiment is often one that seemed least likely at the 
beginning.  The real surprise in this figure is how different models
of physics coming from a very deep level of Nature present distinctly 
different patterns.  These patterns will be visible in 
data that can be collected at the weak interaction scale, data that we 
will gather with the coming generation of high-energy colliders.

\subsection{Two variant phenomenologies}

Before I discuss how we will collect the information 
displayed on the Dine-Nelson plot, I should note other some other 
features of phenomenology which can give insight into the mechanism of 
supersymmetry breaking. In particular, there are sometimes new reactions 
which are specific to a particular mechanism of supersymmetry breaking
whose observation would give evidence of that mechanism.  I will give
two examples here.

Barbieri,  Hall, and Strumia\cite{barbhall}
 have noted that, if the lepton superpartners of the 
three generations receive universal masses at the string scale, the
mass matrix of the 
sleptons grand-unified with the top quark can receive large radiative
corrections proportional to the square of the top quark Yukawa coupling.
These corrections upset the preset universality and can lead to lepton
flavor  violation by flavor mixing  through the third-generation 
sleptons. In Ref. \citen{barbhall}, the consquences of this idea 
are worked out for 
$\mu\to e \gamma$ and other low-energy probes of lepton flavor 
conservation.  In principle, the effect can also be observed directly 
at colliders.  In an $SO(10)$  model which connects the quark and 
lepton mixing angles, one might estimate
\beq
    { \Gamma(\widetilde \tau^- \to  \mu^- \widetilde\chi_1^0) \over
    \Gamma(\widetilde \tau^- \to  \mu^- \widetilde\chi_1^0)}
   \sim \bigl| {m_\tau (A + \mu \tan \beta)\over m_{\tilde\tau}^2} V_{cb}
              \bigr|^2  \sim 10^{-5} \bigl({\tan \beta\over 10}\bigr)^2
\eeq{tauviol}
for $m_{\tilde \tau} \sim A \sim \mu \sim 100$ GeV.  Hall and his collaborators
have shown that that the  effect can be much
larger in other theories of lepton
flavor, and that it might also be visible in other reactions.  For example,
Ref. \citen{fengandhall} discusses models in which there are potentially
observable signals in $\ee\to e^+ \mu^-  \widetilde\chi_1^0 \widetilde\chi_1^0$
and in $e^-e^- \to e^- \mu^- \widetilde\chi_1^0 \widetilde\chi_1^0$.
 
Another interesting possibility arises in model in which supersymmetry is 
broken by gauge theory dynamics at energies relatively close to the weak 
scale.  In any model with spontaneously broken supersymmetry, 
each pair of superpartners couples to the Goldstone fermion of supersymmetry.
For example, assuming that the $\widetilde b$ is an approximate mass
eigenstate, this particles couples to the photon and the goldstino 
($\widetilde g$) by
\beq
      \Delta \L = {\cos\theta_w m_{\tilde b}\over \Lambda^2}
              \bar{\widetilde b} \sigma^{\mu\nu} F_{\mu\nu} \widetilde g \ ,
\eeq{goldstlag}
where $\Lambda$ is the mass scale of symmetry breaking. In supergravity,
the goldstino is eaten by the gravitino, which obtains a mass of order
$\Lambda^2/m_\Pl$.  For values of $\Lambda$ near the unification scale,
the coupling \leqn{goldstlag} is completely irrelevant to particle physics
experiments.  However, as $\Lambda$ comes below 100 TeV, the decay 
$\widetilde b \to \gamma \widetilde g$ can occur inside a collider 
detector, leading to observable reactions with direct photons such as
$\ee \to \gamma\gamma \widetilde g \widetilde g$\cite{yuangrav}
and $\ee, q\bar q \to \widetilde e^+ \widetilde e^- \to 
 \ee \gamma\gamma \widetilde g \widetilde g$.\cite{scottgrav}

The observation of either of these interesting supersymmetry
phenomena would single out particular models from among the many classes
that we are considering here.  In the remainder of this paper, however,
we will consider only the most conservative picture of supersymmetry 
reactions and concentrate on the question of how we can assemble the 
spectral data needed to construct the Dine-Nelson plot.

\section{Experiments at Hadron Colliders}

In the previous section, I have set out a systematic but somewhat
idealized experimental program aimed at establishing the 
mechanism of supersymmetry breaking.  In this program, one must first
set the scale of supersymmetry partner masses by 
measuring the gaugino  masses and testing gaugino universality.  Then one
must identify the scalar states associated with each flavor and helicity of 
quarks and leptons, and we must measure their masses with sufficient
precision to recognize their pattern on the Dine-Nelson plot. 
In this section and the next, I will compare this idealized program with 
the realistic expectations for experiments at future colliders. 

 Though 
it is important to recognize that supersymmetry might be discovered
in next few years at LEP II or at the Tevatron, so that the 
experimental study of 
supersymmetry could begin with the current generation of accelerators,
I will concentrate my discussion on the expectations for the accelerators
of the next generation.  On the side of hadron colliders, I will discuss
studies done for the LHC; on the side of lepton colliders, I will discuss
the expectations for $\ee$ linear colliders.  Some results on supersymmetry
parameter determination specifically directed to the LEP II program have 
been discussed in Refs. \citen{jonandmatt} and \citen{decarlos}.

In this section, then, I will present the aspects of our general 
program which are discussed in simulation studies of supersymmetry
experiments for the LHC.  These studies have,
for the most part, been directed to shorter-term goals than the ones I have
emphasized here, to the first discovery of supersymmetry, rather than
to the systematic experimental pursuit of the new physics.  In fact, it
 should be easy for
the LHC to discover supersymmetry.  The cross section
for gluino pair production in hadronic collisions is an order of magnitude
larger than that for production of a quark of comparable mass, and the 
expected signature of multijet events with large  missing energy is 
striking and characteristic.

To go deeper than the observation of anomalies, however, will
be difficult at hadron colliders.  The reasons for this do not come from
considerations of relative cleanliness and such experimental matters
which are debated between the hadron and lepton physics communities. 
Rather, they come from the specific predictions of supersymmetry 
phenomenology.  The difficulties and the promise of hadron collider
experiments can be made clearer by reviewing some of the techniques which have
been developed to date for obtaining information on the supersymmetry 
spectrum in this environment.

Before beginning this review, I would like to recall the 
expectation, both in this generation of accelerators and the 
next, that hadron and lepton collider experiments should probe 
roughly the same regions of the parameter space of supersymmetry.
The reason for this is that colored superpartners receive large
positive mass enhancements from 
their coupling to gluons and gluinos.  This is most clear in the 
gaugino sector.
I argued in Section 3.1 that gaugino unification should at least be a 
useful guide to the general properties of the supersymmetry spectrum.
According to \leqn{gunif}, the short-distance
gluino mass $m_3$ should be roughly 
three times the mass parameter $m_2$.  To convert to physical mass values,
we must note that $m_2$ is essentially an upper bound to the lightest
chargino mass, while $m_3$ receives a 15\% upward radiative correction when
converted to the  `pole' mass which determines the kinematics of 
gluino production.  Another similarly large correction, 
which may be of either sign, may appear
if the gluino and squark masses differ by a 
large ratio.\cite{glmassD}
Thus, 
\beq
           m(\widetilde g) >  3.5 (m(\widetilde\chi^+_1)^2 - m_W^2)^{1/2} \ .
\eeq{gmassrel}
Thus, a chargino discovery at 80 GeV which might be made at LEP 2 would
correspond to a gluino at 300 GeV which might be discovered at the 
Tevatron.  A linear collider at 1 TeV would be able to search for charginos
up to 500 GeV; the corresponding gluino mass is 1700 GeV, which is 
roughly the search limit of the LHC if 
$m_{\tilde g} \ll m_{\tilde q}$.\cite{atlas}
  Both of these values are
a factor of two  beyond the naturalness limits given in \leqn{natlim}.
In a similar way, the sleptons are expected to be lighter than the
squarks, though the precise relation is more model-dependent. 
Fig. \ref{fig:ClDN}
contains spectra in which the ratio of squark to right-handed slepton masses 
varies from 2 to 6.  Of course, this correspondence does not mean that
the hadron and lepton colliders are competing to discover the same 
information.  In fact, as we will see, quite the reverse is true.

\begin{figure}
 \epsfysize=5.6in
\centerline{\epsffile{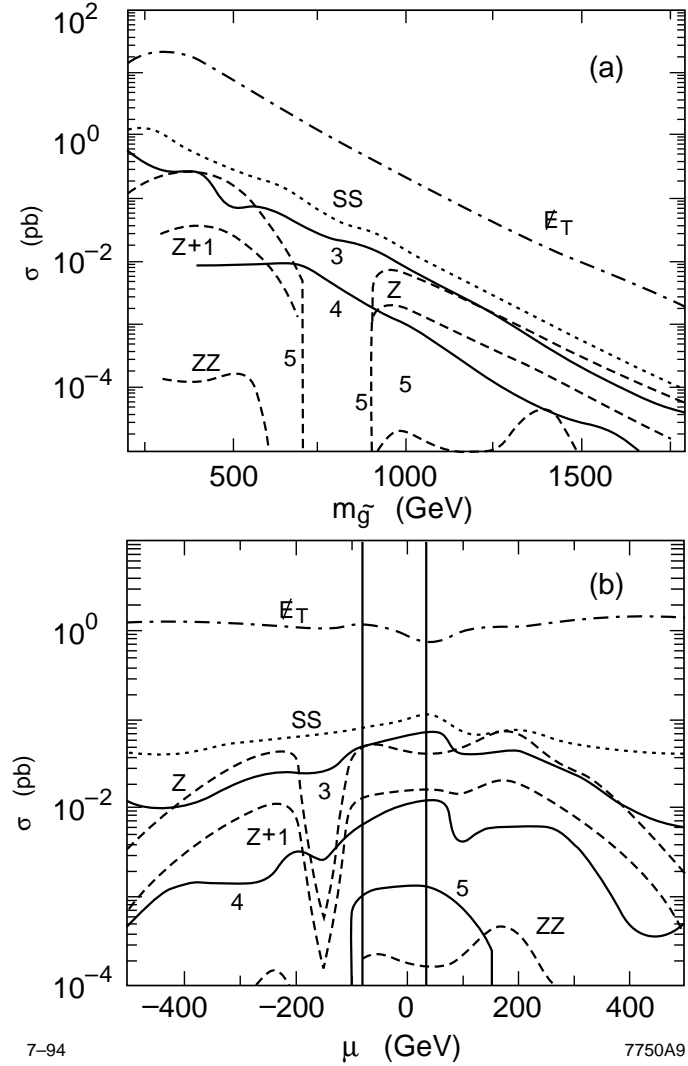}}
\caption{Cross sections for various signatures of 
       supersymmetry observable at the LHC, 
from 
   Ref. \protect\citen{bcascade}.  
The various curves give the cross sections for missing
transverse energy, same-sign dilepton production, multilepton events,
and multilepton + $Z$ events.  The cross sections are shown
 (a) as a function of the mass of the gluino,
       for $m({\widetilde g}) = m({\widetilde q})/2$ and 
         $\mu = -150$ GeV, (b)
       as a function of the parameter $\mu$ for a fixed gluino mass
         equal to  750 GeV.}
\label{fig:LHCsigs}
\end{figure}

Hadron colliders provide many striking signatures of supersymmetry.  The most
basic signature is that of missing energy in multijet events.  But 
the production of supersymmetric particles can also lead to interesting
multilepton and $Z^0$ plus lepton topologies. A summary of event rates
at the LHC
for a variety of increasingly exotic reactions is shown in Fig. 
\ref{fig:LHCsigs}.\cite{bcascade}
These exotic final states arise from decays in which the gluino or
squark which is the primary product of the hadronic reaction decays to 
a neutralino or chargino, which then decays by a cascade to reach the 
lightest superparticle.\cite{cascade}
  An example of such a cascade decay is 
\beqa
   pp \to&\quad  & \widetilde g \to q\bar q + \widetilde \chi_2^+ 
                 \to \ell^+ \nu +  \widetilde \chi_2^0 \to q\bar q + 
                \widetilde \chi_1^0                         \CR
      &+ & \widetilde g  \to
 q \bar q + \widetilde \chi_3^0 \to Z^0 +  \widetilde \chi_1^0
\eeqa{cascade}

 The appearance of these many topologies is a strength
of the hadronic window into supersymmetry, but it is also its weakness.
First, because superpartners are pair-produced, and each partner decays
with missing energy, it is not possible to reconstruct a superpartner
as a mass peak.  The reaction shown in \leqn{cascade} illustrates that
supersymmetry reactions can contain sources of missing energy from 
$\nu$ or $Z^0$ emission in addition to that from the final neutralinos.
Of course, in hadronic collisions, the initial parton energies
and polarizations are also unknown.
Thus, analyses of supersymmetry parameters must be based on overall
hadronic reaction rates, or on other observables which integrate over
the underlying kinematic parameters. To interpret such variables, one 
needs a trustworthy model of the reaction being studied.  But now we come
to the second problem.  The pattern of squark and gluino decays is 
influenced by the spectrum and mixings of charginos and neutralinos and
changes as the parameters of these states move in the $(m_2,\mu,\tan\beta)$
space.   If one relies only on data from hadronic supersymmetry processes,
the dependence on these parameters enters as an essential modelling
ambiguity.

To clarify these issues, I would like to describe a number of methods 
proposed in the literature for the detailed measurement of supersymmetry
parameters.  Before turning to strongly interacting particles, I will 
comment on color neutral states.
Hadronic collisions can also access the chargino and neutralino states
directly, through the reactions
\beq
   q\bar q \to   \widetilde \chi^0  \widetilde \chi^0 , \qquad 
  q\bar q \to\widetilde \chi^0 \widetilde \chi^+ \ .
\eeq{chrxn}
The second of these reactions is a potential source of trilepton events,
and therefore has been discussed as an
 interesting mode for the discovery of supersymmetry
at the Tevatron collider.\cite{trileps}
  Baer and collaborators have noticed that this
reaction can also give some spectral information:  The dilepton 
spectrum in trilepton events falls off sharply at dilepton masses 
equal to the mass difference  $m(\widetilde \chi^0_2)-m(\widetilde \chi^0_1)$,
allowing a measurement of this parameter of the neutralino mass
matrix.\cite{btri}
  Sleptons can also be discovered at hadron colliders.  An analysis
of the slepton signal at LHC, using as the signature acoplanar isolated
leptons, is given in Ref. \citen{bslep}.
This signal unfortunately has a very low rate,
and also sums the contributions of the partners of left- and 
right-handed sleptons, so it is not promising for an accurate mass
determination.

\begin{figure}
 \epsfysize=4.2in
\centerline{\epsffile{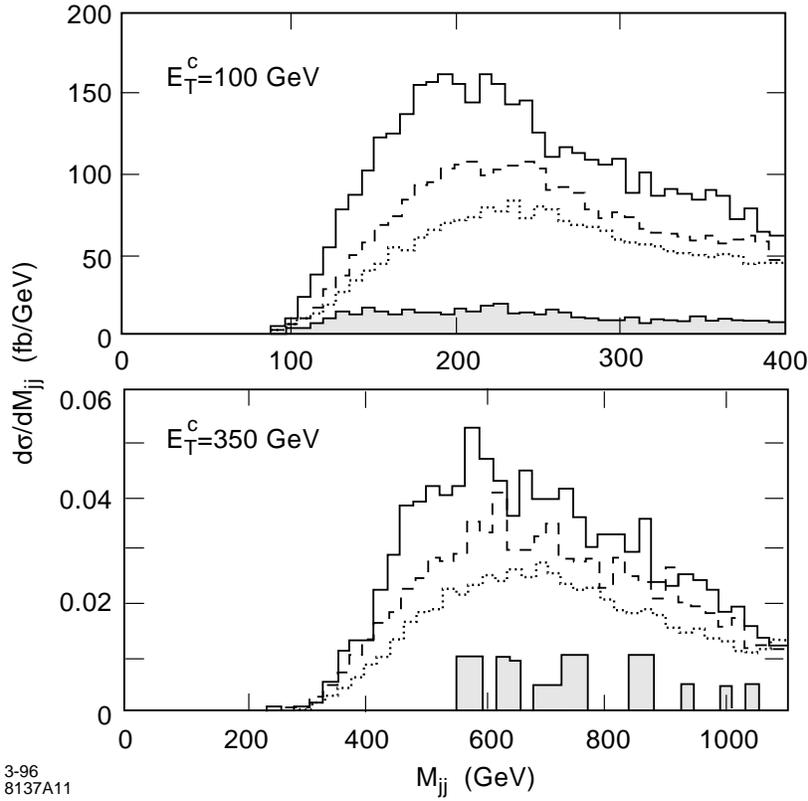}}
\caption{Mass distribution of the estimator of Baer \etal, Ref. 
\protect\citen{bjets},
  for two 
       different ranges of gluino mass: (a) using $E_T^c = 100$ GeV, the
      distributions are shown for $m({\widetilde g})$ = 296, 340, 369 GeV; 
        (b) using $E_T^c = 350$ GeV, the
      distributions are shown for $m({\widetilde g})$ = 773, 885, 966 GeV.
     The simulation assumes that the squarks are much heavier than the 
       gluinos. }          
\label{fig:Tatasigs}
\end{figure}

  For the strongly interacting superpartners, we should hope that the 
hadron colliders can give us accurate mass measurements.  Let us 
consider first the gluino mass measurement.  This is 
simplest if the supersymmetry parameters are such that $m_{\widetilde g}
< m_{\widetilde q}$, and I will restrict my attention to that case for 
 a moment.
There is one proposed estimator for the gluino mass that does peak 
sharply, proposed some time ago by Barnett, Gunion, and Haber.\cite{glmass}
  These 
authors suggested that one should select events with like-sign
dileptons and combine a lepton momentum with the momentum vectors of the
nearest appropriately hard jets.  The resulting mass distribution
roughly tracks the gluino mass 
and has a width of about 15\%.  Baer, Chen, Tata, and 
Paige have criticized this analysis for omitting some backgrounds, 
but have introduced their own 
observable applicable simply to 
missing energy events.\cite{bjets}
  In events with missing transverse energy 
greater than some criterion $E_T^c$, and with two jets in one hemisphere
with transverse energy greater than $E_T^c$, they examine the mass 
distribution of these two jets.  Mass distributions generated by 
Monte Carlo are shown in Fig. \ref{fig:Tatasigs} for sets of three
values of the gluino mass differing by 15\%.   This analysis makes 
plausible that such integral variables can produce a gluino mass
estimate of reasonable accuracy.

\begin{figure}
 \epsfysize=3.0in
\centerline{\epsffile{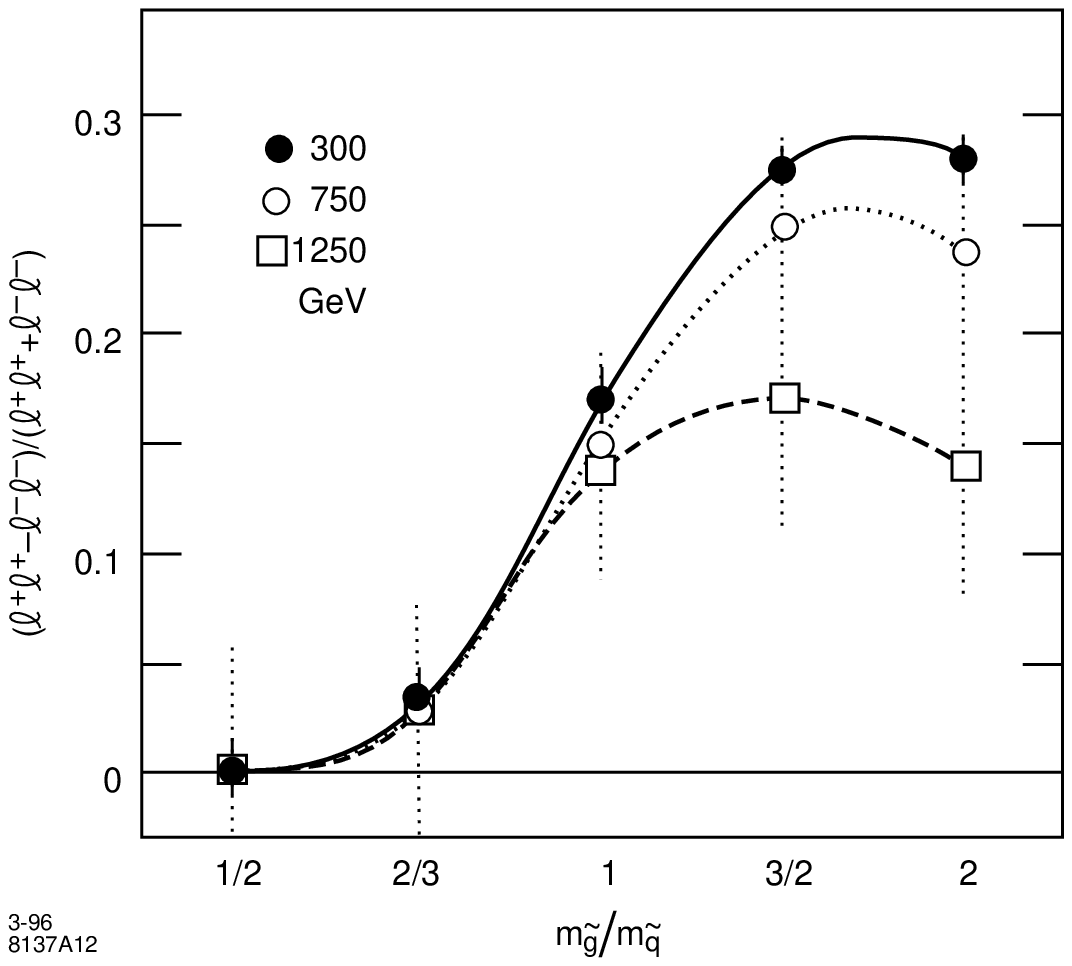}}
\caption{ Dependence of the asymmetry between dilepton
          events with $\ell^+\ell^+$ to those with $\ell^-\ell^-$, as a 
       function of the mass ratio of the squark and gluino, from 
   Ref. \protect\citen{atlas}.
       The three curves represent three different values of the lighter
 of the gluino and quark masses.}
\label{fig:Basa}
\end{figure}

In order to understand whether the gluino is in fact lighter than the 
squarks, and to measure the mass ratio, a number of techniques can be 
employed.  For example, the $\widetilde q_R$ typically  decays dominantly into 
the lightest neutralino, so if these particles are light the missing 
energy signature is stronger and the jet multiplicity is smaller.
The use of jet multiplicity to probe the ratio of the squark and gluino
masses in discussed in Ref. \citen{bjets}.  An additional amusing probe of 
the squark-gluino mass ratio has been studied by Basa\cite{basa} and by the 
ATLAS collaboration.\cite{atlas}  
If squarks and sleptons are comparable in mass,
one of the major processes for supersymmetry production at the LHC 
is 
\beq
              q + q \to  \widetilde q  \widetilde q 
\eeq{sqarkp}
by  $t$-channel gluino exchange.  Since there are more up quarks than
down quarks in the proton, this reaction produces an excess of 
$\ell^+\ell^+$ over $\ell^-\ell^-$ like-sign dilepton events.  The 
asymmetry peaks when the squark and gluino masses are roughly
comparable, as shown in Fig. \ref{fig:Basa}.  On the other hand,
the total rate of
like-sign dilepton events falls as the gluinos become heavier than the 
squarks.  Thus, it is possible at least in principle to determine the mass 
ratio from these two pieces of information.

These observables give the flavor of supersymmetry mass determinations in 
ha\-dron\-ic collisions.  There will be considerable information available,
if one can learn how to use it.  This information resides in integrated
reaction rates for various supersymmetry production processes, and in the 
rates of exotic multilepton reactions.  Unfortunately, the spectral 
pattern is coupled in these observables to the detailed model of squark
and gluino decay, which contains the full complexity of the chargino and
neutralino mixing problem.  

The task of separating these components and
extracting the supersymmetry mass parameters purely from hadronic
cross sections seems like a nightmare.  In fact, none of the analyses 
I have just described have yet been carried out as systematic surveys
over parameter space.  It is not so easy to choose a parameter space of
sufficiently low dimension that it can be surveyed systematically.

Fortunately, the situation looks much brighter if it is assumed that 
there will be an $\ee$ linear collider operating in the same time 
period as the LHC.  I will explain in the next section that experiments
at $\ee$ colliders in provide an array of tools to accurately measure not
only the chargino and neutralino masses but also their mixing angles.
Thus, these experiments will provide the values of the underlying 
supersymmetry parameters needed to explicitly model squark and gluino
decays.  Armed with this model, experimenters at the LHC will be able
to convert their integrated observables into constraints on the spectrum
of squark and gluino masses.  In this way, the extensive energy reach of
the LHC can be used effectively to provide values of the squark and gluino
masses, in a range well beyond the reach of the $\ee$ colliders, to 
accuracies of 10--15\%.

\section{Experiments at $\ee$ Colliders}

Now I will turn to a review of the expectations for supersymmetry 
experiments at the proposed $\ee$ linear colliders which should carry
electron-positron experimentation to the next step in energy.  For 
concreteness, I will have in mind a linear collider with a center-of-mass
energy of 500 GeV and a design luminosity of 50 fb$^{-1}$/yr; this accords
with the current SLAC and KEK designs.  These designs evolve 
smoothly to 1 TeV in the center of mass; the corresponding reach
of almost 500 GeV in the chargino mass covers a region of parameter space
comparable to that
covered by the  LHC, as I have noted above.  A more general review of
the capabilities of $\ee$ linear colliders can be found in Ref. 
\citen{meandHitoshi}.

\subsection{Gaugino masses}

I will begin by discussing $\ee$ experiments on charginos and neutralinos.
In the estimate \leqn{natlim}, I noted that the lighter chargino
and neutralinos are likely
be among the lightest particles of the supersymmetry spectrum.  We can then
use the reactions of these particles to determine the gauginos masses 
$m_1$ and $m_2$, and also to learn more detailed information about the 
values of the general parameters of the supersymmetric model.

\begin{figure}
 \epsfysize=2.2in
\centerline{\epsffile{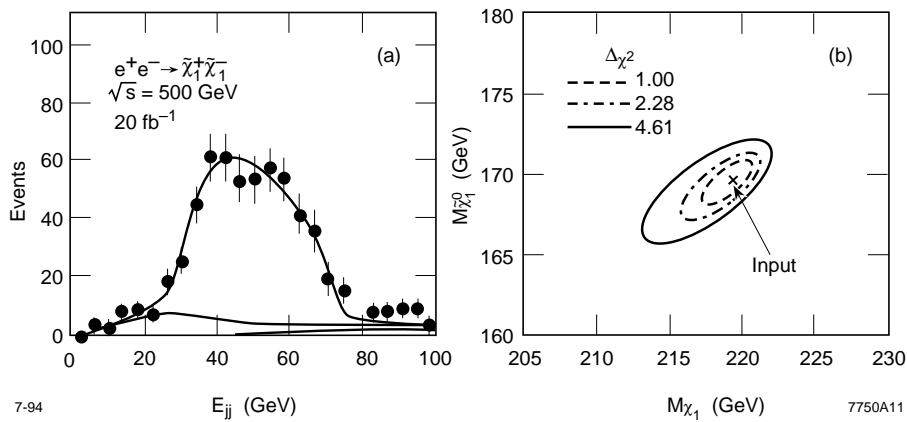}}
\caption{Determination of the lightest chargino mass in the decay 
       $\widetilde\chi_1^+ \to q\bar q \widetilde\chi_1^0$, according
        to the simulation results of Ref. \protect\citen{tsuka}.  The
         right-hand
        figure shows the $\chi^2$ distribution as a function of the 
        masses of $\widetilde\chi_1^+$ and $\widetilde\chi_1^0$.}
\label{fig:chimass}
\end{figure}

If the lighter chargino is lighter than the sleptons, it will decay via
\beq
    \widetilde \chi^+_1 \to \ell^+ \nu \widetilde\chi^0_1 \ , 
\qquad \widetilde \chi^+_1 \to q\bar q \widetilde\chi^0_1 \ .
\eeq{chidecay}
Imagine, then, selecting events with the reaction
$\ee\to \widetilde \chi^+_1 \widetilde \chi^-_1$ in which one 
chargino decays to quarks and the other to leptons.  The invariant mass
of the $q\bar q$ system has a predictable distribution whose kinematic 
endpoints determine the mass of the parent $ \widetilde \chi^+_1$
and the mass of the  $\widetilde \chi^0_1$ into which it decays.
Simulations of the reconstruction of this distribution at future linear
colliders show that these masses can be determined quite accurately.  For
example, in the simulation shown in Fig. \ref{fig:chimass}, these two
masses are determined to 3\% accuracy.

I have stressed in Section 3.1, however, that the determination of the 
masses of  charginos and neutralinos is only the beginning of what is 
needed.  In order to determine the underlying supersymmetry breaking 
parameters of the theory, and to resolve the problem of cascade decays
which 
enters the analysis of the signatures of supersymmetry in $pp$ collisions,
we must also determine the mixing angles which arise when the mass 
matrices are diagonalized.  Fortunately, lepton colliders offer
particular incisive tools which allow one to analyze the 
mixing of chargino and neutralino states.  I will now present two 
techniques for doing this.
In this discussion, I will discuss the 
formulae for $\ee$ cross sections to supersymmetric particle pairs,
but only 
in a rather schematic way.  A very useful compilation of the formulae
for supersymmetry production in $\ee$ reactions can be found in Ref.
\citen{bee}.

\begin{figure}
 \epsfysize=1.3in
\centerline{\epsffile{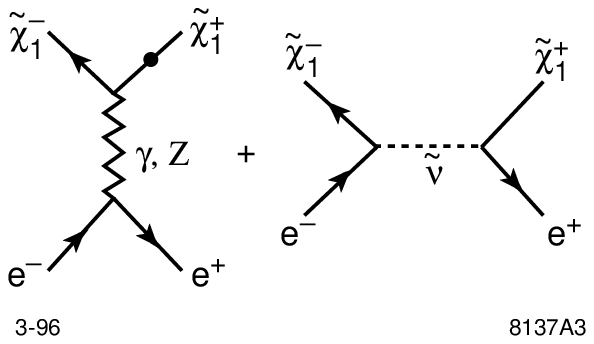}}
\caption{  Diagrams contributing to the process
   $e^- e^+ \to \widetilde \chi^+_1 \widetilde\chi^-_1$.}
\label{fig:ChProd}
\end{figure}

We first consider
the reaction $e^-e^+ \to 
 \widetilde \chi^+_1 \widetilde\chi^-_1$,
making use of  the highly polarized
electron beams which are anticipated for linear collider experiments.
In Ref. \citen{tsuka}, some 
wonderful observations are made about this process.
To understand these, imagine first that we study the reaction at very 
high energy, so high that we can ignore all masses.  Now assume that the 
electron beam can be polarized completely in the right-handed orientation.
Since right-handed electrons do not couple to the $SU(2)$ gauge 
interactions, the second diagram in Fig. \ref{fig:ChProd}
 vanishes.  In addition,
the first diagram in Fig. \ref{fig:ChProd}
involves only the linear combination of $\gamma$ and $Z^0$ which gives 
the $U(1)$ (hypercharge)
 gauge boson.  But the $U(1)$ gauge boson does not couple
to $W$ superpartners.  Thus, this diagram only involves the Higgs
superpartners.  If we project onto the lowest mass eigenstate, the 
rate of the process  $e^-_Re^+ \to 
 \widetilde \chi^+_1 \widetilde\chi^-_1$, will be proportional to the 
squares of the mixing angles linking the $\widetilde h^-_1$ and 
 $\widetilde h^-_2$ to this mass eigenstate.

\begin{figure}
 \epsfysize=3.5in
\centerline{\epsffile{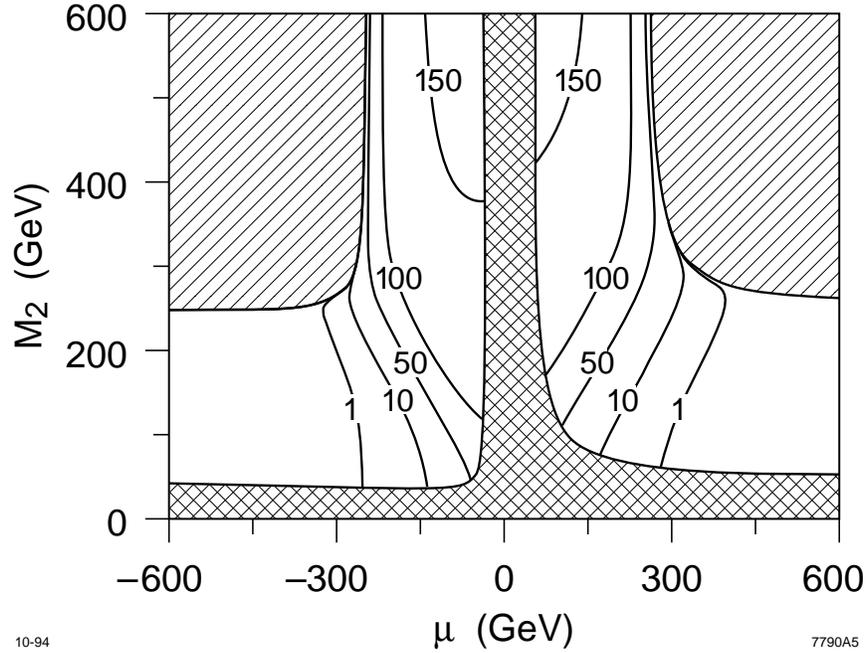}}
\caption{ Total cross section for the process
    $e^-_R e^+ \to \widetilde \chi^+_1 \widetilde\chi^-_1$, in fb, 
as a function
     of $m_2$ and $\mu$, for $\tan\beta =4$, from Ref. 25.
             The selected region is that
      in which the lightest chargino is too heavy to have been discovered
      at the $Z^0$ but is accessible to a 500 GeV $\ee$ collider.}
\label{fig:ChCross}
\end{figure}

 The promise which is 
suggested by this high-energy analysis is actually realized under
more realistic conditions.	In Fig. \ref{fig:ChCross},
 I plot contours of this
polarized cross section for an $\ee$ collider at 500 GeV in the relevant
region of the $(m_2,\mu)$ plane.  You can see that the cross section
maps out this plane, giving the location chosen by Nature,
up to a two-fold ($\mu \leftrightarrow 
-\mu$) ambiguity, for any determined value of the chargino mass.

\begin{figure}
 \epsfysize=3.5in
\centerline{\epsffile{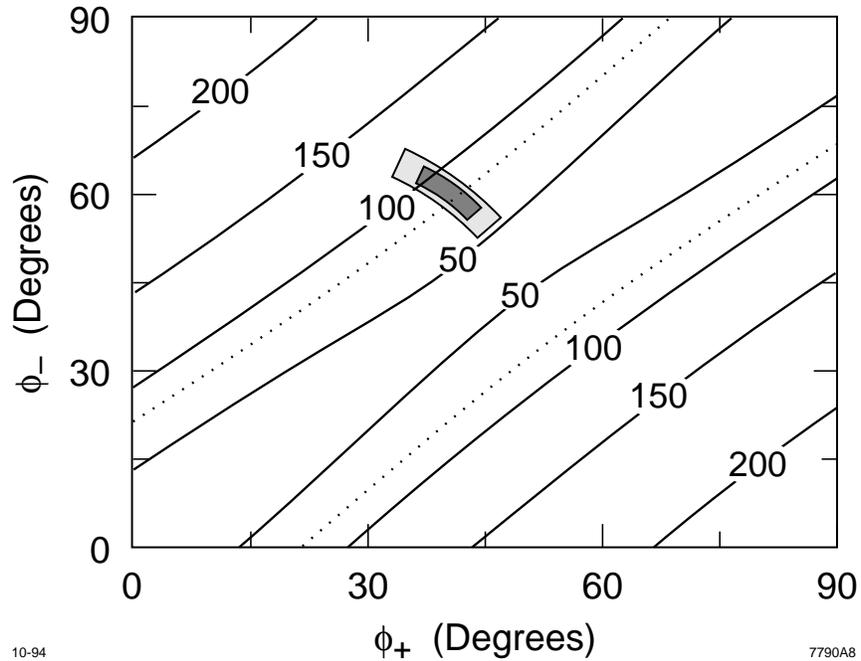}}
\caption{Determination of the two mixing angles of the chargino mass matrix
     from the two chargino masses and the total cross section and 
       forward-backward asymmetry for 
  $e^-_R e^+ \to \widetilde \chi^+_1 \widetilde\chi^-_1$, 
      according to the simulation results of Ref. \protect\citen{feng}.
   The larger box represents the constraint from 30 fb$^{-1}$ of data,
  the smaller box from 100 fb$^{-1}$.}
\label{fig:chiangles}
\end{figure}

Actually the chargino pair production cross section contains even more 
information.  Going back to the limit of very high energies, the
angular distribution for an $e^-_R$ to produce a right-handed fermion
is proportional to $(1+\cos\theta)^2$, while the angular distribution
to produce a left-handed fermion is $(1-\cos\theta)^2$.  Thus, the 
forward production of $\widetilde\chi^-_1$ is given by the mixing angle 
for $\widetilde h^+_2$, while the backward production is controlled by the 
mixing angle for  $\widetilde h^-_1$.  Thus, measurement of both the total
cross section and the forward-backward asymmetry for this process gives
the two mixing angles  needed to diagonalize the chargino mass matrix
\leqn{charginom}.  In this analysis, one must assume that there are  only
two of Higgs doublets that the weak scale (as is required for the grand
unification of couplings), but there are essentially no other
model-dependent assumptions.
The expected constraints on the two mixing angles, for a particular
point studied in the simulations of Ref. \citen{feng}, are shown in 
Fig. \ref{fig:chiangles}.

A second method for determining the gaugino mixing parameters involves
the production of electron partners, selectrons.  There are two 
selectrons, one the partner of $e^-_R$, the other the partner of $e^-_L$.
(These states can be easily distinguished by the polarization 
asymmetry of their production in $\ee$ reactions.) 
 I will discuss the event selection and mass measurement for selectrons
in Section 5.3.  For the moment, we need only note that the selectrons are
expected, in all of the models shown in Fig. \ref{fig:ClDN}, to have masses 
comparable to that of the lightest chargino, so that they should also be
found in the early stages of the experimental program at a linear collider.

\begin{figure}
 \epsfysize=2.5in
\centerline{\epsffile{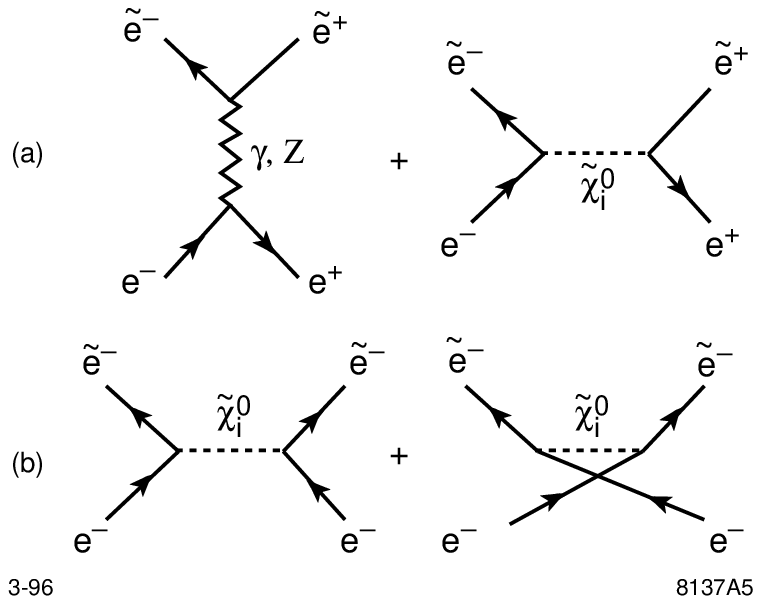}}
\caption{ Diagrams contributing to the processes
  (a)  $e^- e^+ \to \widetilde e^- \widetilde e^+$;
  (b)  $e^- e^- \to \widetilde e^- \widetilde e^-$.}
\label{fig:SeProd}
\end{figure}

The Feynman diagrams which
contribute to selectron production in $\ee$ annihilation are 
shown in Fig. \ref{fig:SeProd}(a).  The second diagram involves neutralino 
exchange.  Although this diagram is exotic, it typically dominates,
since the lightest neutralino is usually lighter than the $Z^0$ and 
the diagram is a $t$-channel rather than an $s$-channel exchange.
A related process is that of selectron production in $e^-e^-$ collisions.
Here the reaction is mediated only by neutralino exchange diagrams, 
in the $t$- and $u$-channel.

To discuss these processes, it is convenient to define `neutralino 
functions', in the following way: Let $V_{ij}$ be the orthogonal matrix
that diagonalizes \leqn{neutralm}, with the first index denoting a 
weak eigenstate and the second denoting a mass eigenstate.  Define
\beqa
       V_{Ri} &=&   -  {1\over \cos\theta_w}V_{1i}\CR
       V_{Li} &=&   -  {1\over 2\cos\theta_w} V_{1i} - {1\over 2\sin\theta_w}
                              V_{2i}  \ .
\eeqa{VLRdefs}
Then define, for $a,b= L,R$,
\beqa
       \N_{ab}(t) &=& \sum_i  V_{ai} {m^2_1\over m^2_i -t} V_{bi} \CR
    \M_{ab}(t) &=& \sum_i  V_{ai} {m_1 m_i\over m^2_i -t} V_{bi} 
\eeqa{NMdefs}
where the sum runs over the four neutralino mass eigenstates, $m_i$ is the
mass of the $i$th neutralino, and $m_1$ has been introduced to make the
functions dimensionless.  The neutralino functions are simply related to the
production cross sections, for example,
\beqa
  {d\sigma\over d\cos\theta}
 (e^-_R e^+& \to &\widetilde e^-_R  \widetilde e^+_R)
           \CR
    = \  & {\pi \alpha^2\over 4s}&
 \bigl[ 1 + {\sstw\over \cstw}{s\over s-m_Z^2} - {s\over m_1^2}
                     \N_{RR}(t) \bigr]^2 \beta^3 \sin^2\theta \ .
\eeqa{samplecross}
The functions $\N_{RR}$, $\M_{RL}$, $\N_{LL}$ enter the formulae for the 
production of $\widetilde e^-_R\widetilde e^+_R$, 
 $\widetilde e^-_R\widetilde e^+_L$, and
 $\widetilde e^-_L\widetilde e^+_L$, respectively, in $\ee$ annhiliation;
the opposite three combinations enter into  the production cross
sections for $e^-e^-$.

\begin{figure}
 \epsfysize=3.25in
\centerline{\epsffile{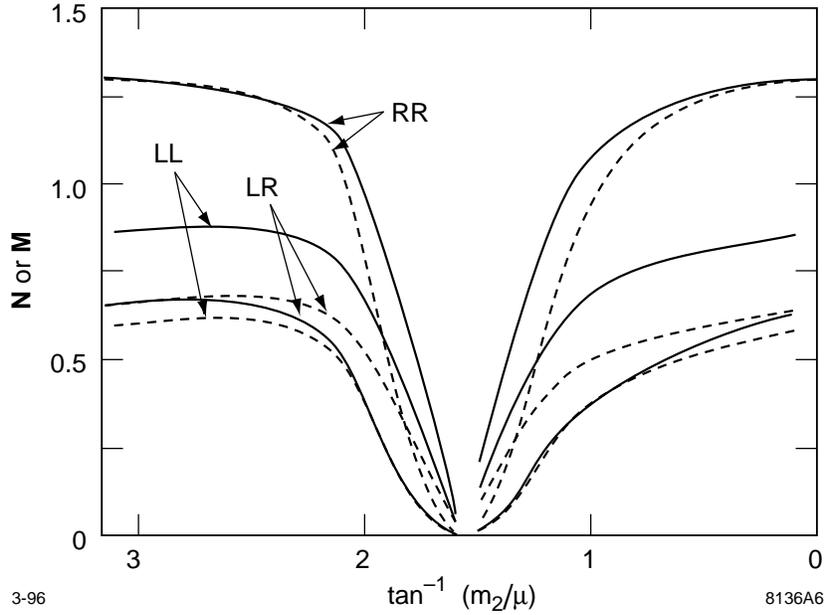}}
\caption{ Values of the `neutralino functions' 
      $\N_{ij}$, $\M_{ij}$, at $t=0$, as a function of the angle
      in the $(m_2,\mu)$ plane: $\alpha = \tan^{-1}(\mu/m_2)$.  The 
      solid curves denote the predictions for selectron production
       in $\ee$ collisions, the dotted curves for selectron production
       in $e^-e^-$ collisions.}
\label{fig:NFcns}
\end{figure}

The neutralino functions are full of information about the neutralino
mixing problem.  As an example, I plot in Fig. \ref{fig:NFcns}
 the values of the six
neutralino functions, extrapolated to $t=0$, along a contour of 
constant chargino mass in the $(m_2,\mu)$ plane.  These variables also
map the position in this plane. Though it is not shown here, the relative
heights of the curves are sensitive to the value of $m_1/m_2$ and thus
provide a test of gaugino unification.  A detailed simulation of 
selectron pair production which uses these ideas to extract
$m_1$, $m_2$ and $\mu$ has been presented in 
Ref. \citen{tsuka}.  The results of that paper, which assume 
50 fb$^{-1}$ of data, correspond to a test of the gaugino unification of
$m_1$ and $m_2$ at the 5\% level.

\subsection{$\tan\beta$}

Unfortunately, there is no method known which systematically determines
$\tan\beta$ throughout its whole range of possible values.  I will 
discuss here four methods, of which the first two apply mainly for 
small or intermediate values of $\tan\beta$ and the last two require probes
at higher energy.  It is likely that some combination of these methods can 
determine $\tan\beta$ well enough to interpret experimental data on the 
superparticle spectrum.

The first method for  determining $\tan\beta$ goes back to the chargino
production cross section discussed in Section 5.1.  I argued there that
it is possible to determine the mixing angles needed to diagonalize the
chargino mass matrix; from these, one can deduce the off-diagonal elements
of this mass matrix.  But note from \leqn{charginom} that the ratio of these
elements is just equal to $\tan\beta$.  Since these off-diagonal elements are
related by supersymmetry to the vertices which give mass to the $W$ boson,
this ratio is model-independent.  In Ref. \citen{feng}, it 
was remarked that the determination of the 
chargino mass matrix discussed there could be interpreted as a $\tan\beta$
measurement.  Then this parameter could be determined with an accuracy of 
was  3\% at $\tan\beta = 4$, for a parameter set in 
which the lightest chargino was a 
roughly  equal mixture of gaugino and Higgsino.

A second method for determining $\tan\beta$ has been proposed by 
Nojiri.\cite{nojiri}
This involves a beautiful supersymmetry observable for linear colliders,
the polarization of the $\tau$ leptons produced in $\widetilde\tau$ decay.
The $\tau$ polarization is now known to be straightforwardly measurable in
$\ee$ experiments.  The polarization of $\tau$'s from $\widetilde\tau$ 
decay contains information on the mixing of the two $\widetilde\tau$ 
eigenstates and on the decay pattern.  For a full discussion of the 
extraction of this information, see Ref. \citen{nojiri}. 
 For the purpose of this 
discussion, I will simply point out that the mixing of 
$\widetilde\tau_L$ and $\widetilde\tau_R$ can be determined from the 
$\widetilde\tau$ cross sections and polarization asymmetry. In the 
following discussion, I will assume for simplicity that the lightest
$\tau$ partner is an unmixed $\widetilde\tau_R$.

\begin{figure}
 \epsfysize=1.0in
\centerline{\epsffile{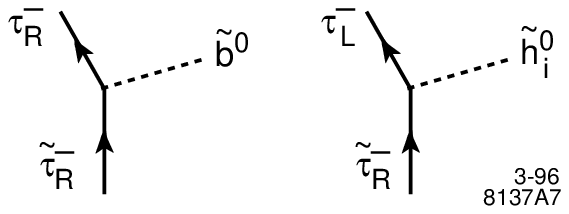}}
\caption{ Components of the decay $\widetilde\tau_R \to 
             \tau \widetilde \chi^0_1$.}
\label{fig:TauC}
\end{figure}

The dominant decay of this scalar should be to $\tau \widetilde\chi^0_1$.
In terms of weak-interaction eigenstates, there are
two amplitudes that contribute to this decay; these are shown
in Fig. \ref{fig:TauC}.
  On one hand, the $\widetilde\tau^-_R$ can decay to a $\tau^-_R$
with the emission of a $\widetilde b^0$.  On the other hand, the  
 $\widetilde\tau^-_R$ can decay to a $\tau^-_L$
with the emission of a $\widetilde h_1^0$.  These two processes give rise
to a nontrivial $\tau$ polarization, given to first order by
\beq
  P(\tau^-) = 1 - {\cstw\over \sstw} {m_\tau^2\over m_W^2}{1\over \cos^2\beta}
             {p(\widetilde h_1^0)\over p(\widetilde b^0)} \ ,
\eeq{taupol}
where $p(\widetilde h_1^0)$ and $p(\widetilde b^0)$ are the probabilities
that the lightest neutralino appears as one of these states.
If we know the content of the lightest neutralino
mass eigenstate in terms of weak eigenstates---and I have 
given methods for determining this in the previous section---this formula
can be solved for $\cos\beta$.  This technique should give $\tan\beta$
measurements below the 10\% level even when the Higgsino component of the
lightest neutralino is rather small.

Ideally, $\tan\beta$ can be determined from the branching ratios of the 
heavy Higgs bosons of supersymmetry.  If the $A^0$ boson of the Higgs sector
has a mass well above the $Z^0$ mass, the lightest Higgs boson $h^0$
has branching ratios close to those of the Higgs boson of the minimal
standard model.  However, the heavy Higgs bosons $H^0$ and $A^0$ have
couplings which reflect the ratio of the two Higgs vacuum expectation 
values.  For example,
\beq
       {\Gamma(H^0\to t\bar t)\over \Gamma(H^0 \to b\bar b)} = 
        \bigl({m_t\over m_b}\cot^2\beta \bigr)^2
       \bigl( 1 - {4m_t^2\over m_H^2}\bigr)^{1/2} \ .
\eeq{Higgsrat}
Unfortunately, these heavy Higgs bosons have masses of order 500 GeV in 
typical models, and they must be pair-produced (except in $\gamma\gamma$
collisions); thus, they may be difficult to find in the early stages of
linear collider experimentation.

As a last resort, there is one interesting determination of $\tan\beta$
that will be available from the LHC.  The processes $H^0, A^0 \to 
\tau^+\tau^-$ is a possible mode for observing the heavy Higgs bosons at the
LHC, but only if the branching ratios to $\tau^+\tau^-$ are enhanced
by a large value of $\tan\beta$.\cite{cms,atlas}
  If this signature can be observed, then
$\tan\beta > 10$, which is already sufficient information to evaluate
the scalar mass contribution \leqn{Dterm} to a reasonable accuracy.

\subsection{Scalar partner masses}

Once we have determined the values of $m_2$ and $\tan\beta$ and have 
either verified gaugino universality or measured the independent gaugino 
masses, we are ready to measure the masses of squarks and sleptons that 
will contribute to the Dine-Nelson plot.  From the various spectra shown
in Fig. \ref{fig:ClDN}, it is likely that the sleptons can be found at least
at a 1 TeV linear collider, and it is possible that the squarks could also
be found at such a facility.

\begin{figure}
 \epsfysize=2.1in
\centerline{\epsffile{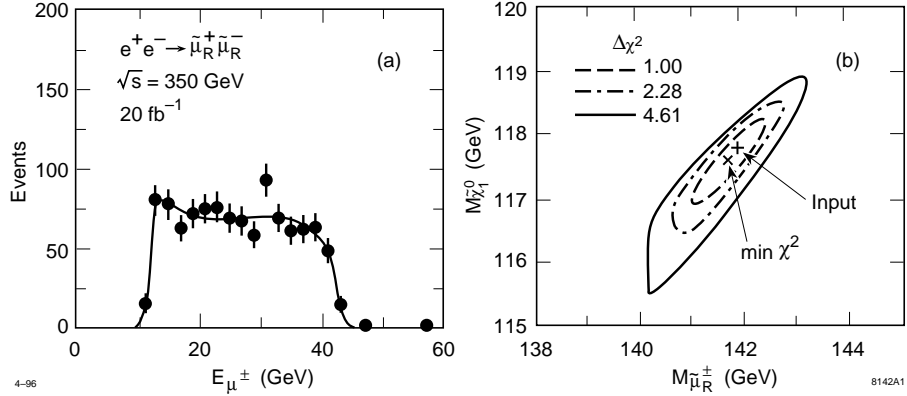}}
\caption{Determination of the $\widetilde\mu_R$ mass in the decay 
       $\widetilde\mu_R^- \to \mu^- \widetilde\chi_1^0$, according
        to the simulation results of Ref. \protect\citen{tsuka}.  
     The analysis assumes right-handed electron polarization 
            $P = 0.95$.  The right-hand
        figure shows the $\chi^2$ distribution as a function of the 
        masses of $\widetilde\mu_R^-$ and $\widetilde\chi_1^0$.}
\label{fig:mumass}
\end{figure}

Consider first the sleptons.  The simplest possible decay of a slepton
is 
\beq
        \widetilde \ell^- \to \ell^- \widetilde \chi^0_1 \ ,
\eeq{sleptondec}
and this mode has a substantial branching ratio thorough most of the 
parameter space.  This leads to events of the form
$\ee\to \ell^+\ell^- \widetilde \chi^0_1
 \widetilde \chi^0_1$ which are very simple to analyze. The only important
background to these events comes from $\ee\to W^+W^-$, and this can be 
reduced by using a polarized $e^-_R$ beam.  The lepton energy 
distribution from the decay of the scalar $\widetilde\ell$ is flat, with 
a sharp cutoff at the kinematic endpoints.  By fitting the endpoints, one
determines the mass of the parent $\widetilde\ell$ and the mass of the 
$\widetilde \chi^0_1$ into which it decays.  Simulation results on the
determination of the $\widetilde \mu$
mass, taken from Ref. \citen{tsuka}, are shown in Fig.~\ref{fig:mumass}.
This analysis corresponds to a 1\% mass determination for the slepton and
for the lighest neutralino.  The superpartners of left- and right-handed
leptons can be distinguished by their $SU(2)\times U(1)$ quantum numbers as
determined from the values of their production cross sections and 
polarization asymmetries.  Thus, for the sleptons, we can expect to obtain
very accurate measurements to fill in the pattern on the left-hand side
of the Dine-Nelson plot.

\begin{figure}
 \epsfysize=2.5in
\centerline{\epsffile{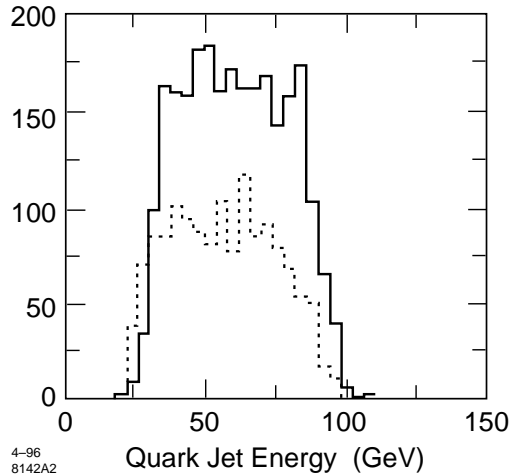}}
\caption{Distribution of quark jet energies in squark decay in one
     region of parameter space studied in Ref. \protect\citen{fandfin}.
    The solid (dashed) histogram refers to events with $e^-_L$
     ($e^-_R$) polarized beams.}
\label{fig:FFjets}
\end{figure}

If squarks lie above the energy range of the linear collider, we must be
content with the 10--15\% measurement of the average squark mass that will
emerge from the analysis of supersymmetry observables at the LHC.  However,
if the linear collider can produce squarks, some more elegant experiments are
possible.  For example, $\ee$ annhilation with an $e^-_L$ beam dominantly 
produces the superpartners of left-handed quarks, while the use of an
$e^-_R$ beam dominantly produces the superpartners of right-handed quarks.
 Feng and Finnell\cite{fandfin} have shown that it is possible to
use this asymmetry to measure the mass difference between $\widetilde q_L$
and $\widetilde u_R$, $\widetilde d_R$.  Consider, for example, the 
particularly favorable region in which all species of squarks decay 
dominantly to $q \widetilde\chi^0_1$.  A simulation of the 
distribution of quark jet
energies from squark production, for a particular point in this region,
is shown in Fig. \ref{fig:FFjets}.  As with the lepton energy distribution
in slepton decay, the quark energy distrbution is flat and cuts off 
sharply at the endpoint.  By comparing the location of the endpoint for 
the two electron beam polarizations, it is possible to measure the 
left-right squark mass difference to 1\% of the squark 
mass.  Using more sophisticated techniques described in Ref. \citen{fandfin},
it is possible to reach comparable precision in other regions of the
parameter space.

One interesting unsolved problem is the question of  how to measure
the  $\widetilde u_R$--$\widetilde d_R$ mass difference.  With this 
quantity under control, the full pattern on the Dine-Nelson plot would
be revealed experimentally.  However, you can see from Fig. \ref{fig:ClDN}
that the precision flavor- and helicity-selected measurement of slepton
masses, plus a reasonable knowledge of the average squark mass, already
distinguishes the various classes of models shown.  This information can 
realistically be made available by combining the results of LHC and linear
collider experiments.

\section{Conclusion}

In this article, I have tried to sketch out the experimental program that
would follow from the discovery of supersymmetry at the weak interaction
scale.  It is an important question whether supersymmetry is present at the 
TeV scale, and whether it is the mechanism of electroweak symmetry 
breaking.  But, if indeed Nature chooses this mechanism, what we have to 
learn at the next generation of colliders goes far beyond this single 
piece of information.  The spectrum of supersymmetric particles contains
information which bears directly on the physics of very short distances,
perhaps even down to the 
unification or gravitational scale.  The challenge will be
to extract this information and study its lessons.

Pursuing this goal, I have set out a three-step program  to clarify the 
physics of the supersymmetry mass spectrum.  To set the scale of 
superpartner masses, we first need to measure the gaugino masses and
the Higgs sector parameter $\tan\beta$. In the process, we must test
the hypothesis of gaugino unversality.  Then, incorporating all of this
information, we can measure the slepton and squark masses and try to 
recognize their pattern as characteristic of a specific messenger of 
supersymmetry breaking.

Electron-positron colliders have a major role to play in this program.
Using their access to the simplest supersymmetry reactions and
the control of beam polarization, these facilities allow
model-independent measurements of the uncolored gaugino masses.
They can also provide accurate, helicity-specific measurements of the 
slepton masses.  If the squarks are sufficiently light, linear  colliders
can be used to measure the helicity-specific squark masses as well.

Hadron collider experiments can be expected to pin down the masses of the 
heavier states of supersymmetry, the squarks and gluinos.  However, the
observables which are useful for hadron experiments require for their
interpretation a detailed model of the 
decay pattern of strongly-interacting superpartners.  Thus,
the interpretation of 
experimental results from hadron collider will also rely on the 
precision information available from $\ee$ colldiers.

It is daunting that the detailed study of supersymmetry will require 
a number of new and expensive experimental facilities.  But we can already
anticipate that these facilities will suffice to give us concrete 
information on the spectrum of superpartners, information we can use to 
determine the mechanism of supersymmetry breaking and the linkage of 
weak-scale supersymmetry to deep theoretical speculations. It is a pleasant
dream that in the future
we might have direct experimental information on the 
physics of the unification or the superstring scale. 
With these new tools---the LHC and the $\ee$ linear collider---we can make
this dream a reality.

\newpage

\section*{Acknowledgements}

The ideas expressed here have been strongly influenced by discussions with
 Howard Baer, Tim Barklow,
Michael Dine, Savas Dimopoulos, Lance Dixon, Keisuke Fujii,
Howard Haber, Jonathan Feng, Gordon Kane,
Hitoshi Murayama, and Xerxes Tata.  I am 
grateful to them, and to many other colleagues at SLAC and in the broader
community, for educating me in this 
area.  I am deeply grateful to Taichiro Kugo and the organizers of the 
YKIS '95 meeting for their hospitality in Kyoto.
 This work was supported by the
Department of Energy under
contract DE--AC03--76SF00515.

\def\Journal#1#2#3#4{#1{\bf #2} (#4) #3}
\def\NPB{Nucl.~Phys.\ {\bf B}}
\def\PLB{Phys.~Lett.\ {\bf B}}
\def\PRL{Phys.~Rev.~Lett.\ {}}
\def\PRD{Phys.~Rev.\ {\bf D}}
\def\ZPC{Z.~Phys.\ {\bf C}}

\end{document}